\documentclass[aps,prl,superscriptaddress,twocolumn]{revtex4-2}
\usepackage{xcolor}
\usepackage{xr}
\usepackage{hyperref}
\usepackage{amsmath,bm,amsfonts}
\usepackage{graphicx}
\usepackage{mathtools}
\allowdisplaybreaks

\makeatletter
\let\@old@ddots\ddots
\def\ddots{\mathclap{\smash{\@old@ddots}}}
\makeatother

\begin{document}

\title{Merons, hedgehogs and magnetoelectric switching in spiral multiferroics}

\author{Luca Maranzana}
\affiliation{Quantum Materials Theory, Italian Institute of Technology, Via Morego 30, Genoa, Italy}
\affiliation{Department of Physics, University of Genoa, Via Dodecaneso 33, Genoa, Italy}

\author{Naoto Nagaosa}
\affiliation{RIKEN Center for Emergent Matter Science (CEMS), Wako, Saitama 351-0198, Japan}
\affiliation{Fundamental Quantum Science Program, TRIP Headquarters, RIKEN, Wako 351-0198, Japan}

\author{Sergey Artyukhin}
\email[]{sergey.artyukhin@iit.it}
\affiliation{Quantum Materials Theory, Italian Institute of Technology, Via Morego 30, Genoa, Italy}

\begin{abstract}
In spiral multiferroics, magnetism induces ferroelectricity, thus holding a promise for novel memory devices where an electric field switches magnetic bits. However, such a switching process, in which magnetic domain walls are moved electrically, is still poorly understood. We find multiferroic domain walls containing arrays of meron (half-skyrmion) strings with a plethora of topological defects, which profoundly affect wall dynamics. Minimum energy walls have alternating meron topological charges and move as relativistic massive particles, with velocity limited by the magnon speed. During domain nucleation, walls with non-alternating meron charges appear. Such defects result in a peculiar non-local dynamics where all the spins in the system rotate, and the wall mobility is suppressed. Meron strings possess 0D hedgehog defects, analogous to Bloch points, that pin the wall to the lattice. This fascinating interplay of magnetoelectric switching with a variety of topological defects and non-local spin dynamics opens a new playground for the electric manipulation of spins.
\end{abstract}

\maketitle
\flushbottom

\paragraph{Introduction. ---}
Domain wall (DW) motion in magnetic materials is a subject of both fundamental and technological interest \cite{Schryer74, Thiele73, Tretiakov08, Clarke08, Foggetti22, Duine10}.
Indeed, switching between two states of a magnetic bit proceeds through the nucleation, propagation and annihilation of DWs. The existing magnetic storage devices use a magnetic field to record (i.e., drive the DWs), leading to considerable power consumption and limiting information density. Multiferroic materials combine multiple coexisting orders and enable cross-control, particularly the electric control of magnetization \cite{Kimura03, Katsura05, Mostovoy06, Sergienko06, Cheong07}. This functionality allows to drive magnetic DWs by an electric field for a drastically more efficient switching process \cite{Roy14}.

\begin{figure}[b]
\includegraphics[width=1\linewidth]{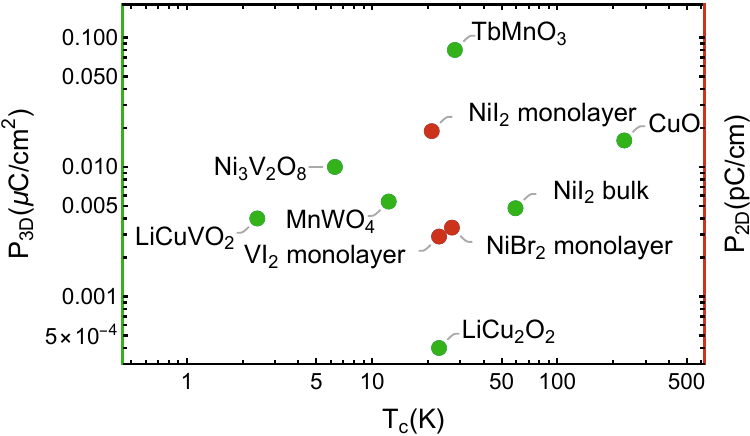}
\caption{\label{fig:Materials}Examples of centrosymmetric spiral multiferroics. $T_c$ denotes the spiral ordering temperature, $P_\mathrm{3D}$ and $P_\mathrm{2D}$ indicate the ferroelectric polarizations for bulk (green) and monolayer (red) materials, respectively.
}\end{figure}

\begin{figure*}[t]
\includegraphics[width=1\linewidth]{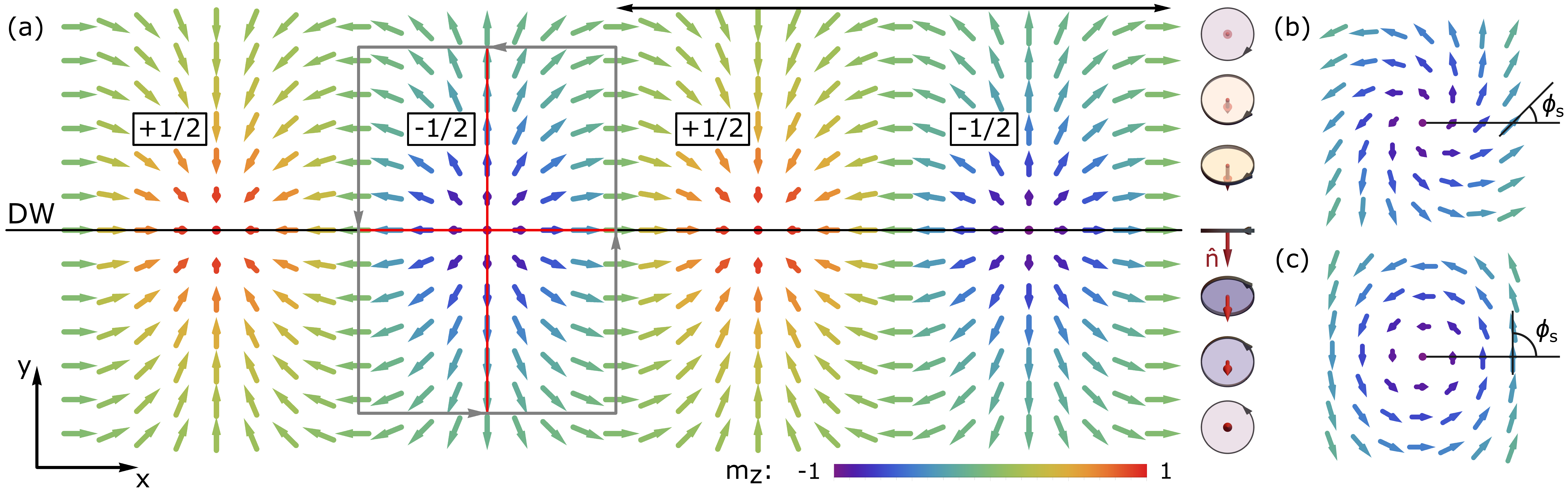}
\caption{\label{fig:StaticDW}(a) Magnetization texture in the $xy$-plane associated with a meron DW parallel to the $xz$-plane (horizontal line). The texture is the same for every fixed $z$. Colors encode the magnetization component $m_z$. The upper domain is a clockwise spiral ($\chi = -1$) with the wave vector along $x$. Across the meron DW, the spin rotation axis $\hat{n}$ rotates by 180$^\circ$ (right inset), leading to a counterclockwise spiral ($\chi = 1$) in the lower domain. The gray contour encircles a meron with topological charge $\mathcal{Q} = -1/2$. In each spiral period, there are two merons with opposite topological charges. The perpendicular red lines show the two mirrors $m'_{x,y}$ of merons. (b,~c) Merons with helicity angles $\phi_s = \pi/4$ (b) and $\phi_s = \pi/2$ (c). The meron encircled in panel (a) has $\phi_s=0$. For $\phi_s \neq 0$, merons still have two discrete symmetries. However, these also involve a rotation in the $xy$-plane. We observe that $\phi_s$ determines the rotation axis of $\hat{n}$. For $\phi_s=0$, $\phi_s = \pi/4$ and $\phi_s = \pi/2$, $\hat{n}$ rotates around $\hat{x}$, $(\hat{x} + \hat{y})/\sqrt{2}$ and $\hat{y}$, respectively.
}\end{figure*}

The most basic non-collinear magnets, spiral magnets, are also prototypical magnetoelectric multiferroics. The magnetic spiral breaks inversion symmetry and induces a ferroelectric polarization ${\bf P}\propto [\hat{n}\times{\bf Q}]$, where $\hat{n}$ is the unit normal to the spin rotation plane and $\bf{Q}$ is the spiral wave vector \cite{Mostovoy06}. $\bf{P}$ is non-zero for a cycloidal spiral, for which spins rotate in a plane that contains $\bf{Q}$. The DW between two spiral domains with opposite chirality consists of an array of vortex or meron (half-skyrmion \cite{Polyakov75, Bogdanov94, Mostovoy23}) strings (see Fig.~\ref{fig:StaticDW}), except for a few wall orientations \cite{Li12}. The existing literature focuses on (vortex-free) Hubert DWs \cite{Hubert98, Nattermann14, Foggetti22} and on vortex DWs \cite{Li12, Nattermann14, Roostaei14, Roostaei15}. Meron DWs, which arise when the anisotropy is weaker, are still widely unexplored. Nevertheless, most of the experimental realizations of spiral magnets (e.g. CuO, MnWO$_4$, Ni$_3$V$_2$O$_8$, and LiCu$_2$O$_2$ \cite{Tokura10}, see Fig.~\ref{fig:Materials}) are in this weak anisotropy regime. In particular, electric field-induced switching has been experimentally studied in MnWO$_4$ \cite{Hoffmann11, Hoffmann13}. The results suggest that the spiral plane rotates across the wall, consistent with meron DWs (see Fig.~\ref{fig:StaticDW}) but not with vortex DWs. Moreover, Hubert DWs have only one possible orientation (i.e. perpendicular to the wave vector). Since two preferred wall orientations are observed, at least one corresponds to meron DWs. Eventually, we expect this DW type to be relevant also in the novel 2D spiral multiferroics (e.g. NiI$_2$, NiBr$_2$, and VI$_2$ \cite{Song22, Sødequist23}, see Fig.~\ref{fig:Materials}).

\begin{figure*}[t]
\includegraphics[width=1\textwidth]{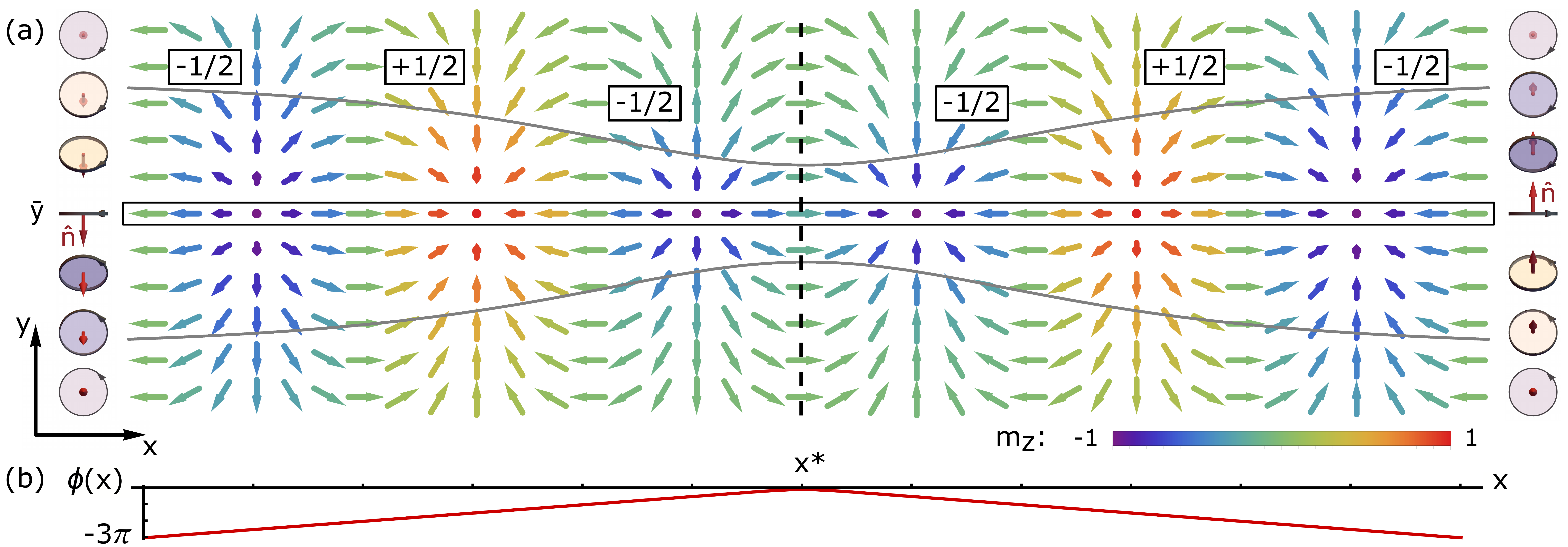}
\caption{\label{fig:Defect}(a) Cross-section in the $xy$-plane of a meron DW containing a defect. Topological charges are not alternating but have signature $\{...-+--+-...\}$. Hence, the total topological charge is $\mathcal{Q}_\mathrm{tot} = -1$. The defect at $x = x^*$ and $y=\bar{y}$ separates two parts of the DW where the spin rotation axis $\hat{n}$ rotates around $\hat{x}$ and $-\hat{x}$ (see side insets). The gray lines show the DW width $\lambda$ as a function of $x$. Near the defect, the DW is narrower. For $y = \bar{y}$ (black box), panel (b) represents the spiral phase $\phi(x)$ so that ${\bf m}(x,\bar{y})=\left(\cos\phi(x),0,\sin\phi(x)\right)$. A Hubert DW at $x = x^*$ separates the asymptotic regions $\partial_x \phi(x) = Q$ and $\partial_x \phi(x) = -Q$.
}\end{figure*}

\begin{figure*}[t]
\centering
\includegraphics[width=1\linewidth]{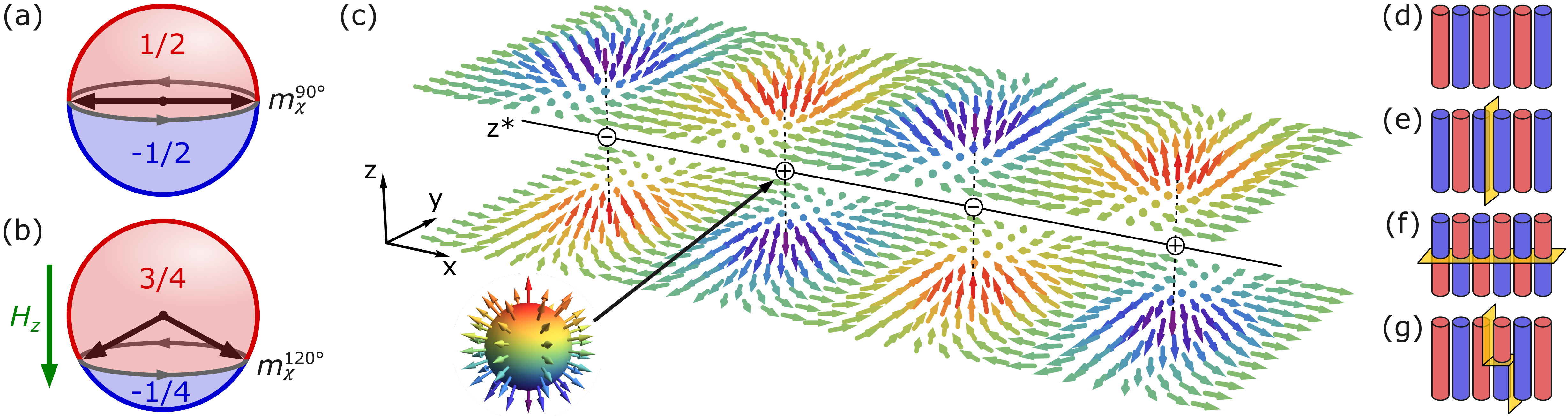}
\caption{\label{fig:TopControl}(a) Region of the sphere wrapped by a positive (red) and a negative (blue) meron for a DW separating two flat spiral domains. The magnetization in the domains map on the equator, resulting in the opposite meron topological charges $\pm 1/2$. (b) An out-of-plane magnetic field $H_z$ stabilizes the conical spiral with a conical angle $\vartheta_c = 120^\circ$, leading to an imbalance between positive and negative merons. (c) Array of hedgehog defects with coordinates ${\bf x}_j = (x_0 + j \pi/Q, \bar{y}, z^*)$ and alternating magnetic monopole charges connecting two half-spaces above and below $z=z^*$ where the meron topological charges have opposite signs. (d--g) Sketches of the meron strings (defect lines indicated in yellow): (d) minimum energy DW (see Fig.~\ref{fig:StaticDW}); (e) defect between two meron strings (see Fig.~\ref{fig:Defect}); (f) array of point defects within the meron strings, corresponding to (c); (g) point defect within a single meron string. In the latter case, defect lines must also be present between the strings.
}\end{figure*}

We analyze the structure and the electric field-driven dynamics of meron DWs in cycloidal spiral magnets by combining the Ginzburg-Landau approach, the collective coordinates method \cite{Tretiakov08, Clarke08}, and atomistic spin dynamics simulations \cite{Landau35, Gilbert04, Skubic08}. The dynamics of a minimum energy wall, for which the meron topological charges alternate, reduces to that of a ``relativistic'' massive particle, where the maximum magnon velocity plays the role of the speed of light. Low-energy defects, corresponding to neighboring merons with the same charges (see Fig.~\ref{fig:Defect}), give rise to DWs with a net topological charge and a characteristic ferroelectric polarization profile, which can be exploited for their detection. These peculiar DWs present non-local dynamics that involve rotations of spins in the entire domains, in contrast to the usual DW motion \cite{Schryer74, Thiele73, Tretiakov08, Clarke08}, where spins rotate only within the wall. Systems in which spiral layers are coupled antiferromagnetically (e.g. TbMnO$_3$) represent an exception to such non-local dynamics due to the mutual cancellations of topological charges in neighboring layers. We also find meron DWs to host hedgehog point defects (see Fig.~\ref{fig:TopControl}(c--g)) that separate the segments of a meron string with opposite topological charges, carry a magnetic monopole charge, and enhance wall pinning. We validate the results with atomistic spin dynamics simulations using a simplified model with competing nearest and next-nearest neighbor interactions, as well as a realistic model for TbMnO$_3$ \cite{Mochizuki09a, Mochizuki09b, Mochizuki10a, Mochizuki10b, Mochizuki10c, Mochizuki11a, Mochizuki11b, Matsubara15}.

\paragraph{Meron domain wall structure. ---}
To describe a general centrosymmetric spiral multiferroic, we consider the following Ginzburg-Landau Hamiltonian density \cite{Li12}
\begin{eqnarray}
\mathcal{H} &=& J_x \left[-Q^2 (\partial_x {\bf m})^2 + \frac{1}{2} (\partial_{x}^2 {\bf m})^2\right] \nonumber\\
& & + \frac{1}{2} \lvert J_\perp \rvert \left[(\partial_y {\bf m})^2 + (\partial_z {\bf m})^2\right] + K_z m_z^2, \label{eq:H0}
\end{eqnarray}
where the unit vector ${\bf m}({\bf r})$ is the magnetization direction and the lattice constant is set to $1$. The term proportional to $J_x > 0$ promotes a spatial rotation of ${\bf m}({\bf r})$ with wave vector $\pm Q \hat{x}$. $J_\perp < 0$ describes ferromagnetic interactions in the $yz$-planes. The antiferromagnetic case is treated in the final section. $K_z > 0$ corresponds to an easy $xy$-plane anisotropy. In Supplementary Information (SI), we study the connection between these parameters and those of a microscopic model, using TbMnO$_3$ as an example.

The minimum energy spin configurations consist of two spirals, where ${\bf m}(x)$ rotates in the $xy$-plane with chirality $\chi = \pm 1$ (i.e. the spin rotation axis is $\hat{n} = \chi \hat{z}$),
\begin{equation}
\label{eq:Cycloids}
{\bf m}_\chi(x) = \left(\cos(Qx + \phi), \chi \sin(Qx + \phi), 0\right).
\end{equation}
The constant phase $\phi$ corresponds to a translation of the whole spiral in the $x$-direction.


In the following, we consider the DW at $y=\bar{y}$ parallel to the $xz$-plane so that ${\bf m}_\chi(x)$ rotates about $\hat{n} = \hat{z}$ below $\bar{y}$ and about $\hat{n} = -\hat{z}$ above. This DW separates the two spiral domains \eqref{eq:Cycloids} with chirality $\chi = \pm1$. For simplicity, we neglect spiral distortions due to the anisotropy, which is justified when $J_x$ dominates, i.e. $K_z \ll J_x$.
To the first approximation, the lowest energy magnetization texture minimizes the dominant term of the Hamiltonian density \eqref{eq:H0}. Consequently, this configuration is a spiral with wave vector $Q$ for every fixed $y$. While moving across the wall, the spin rotation axis $\hat{n}$ rotates from $\hat{z}$ to $-\hat{z}$ (see Fig.~\ref{fig:StaticDW}). Thus, the DW magnetization texture takes the form
\begin{equation}
\label{eq:Ansatz}
{\bf m}(x,y) = \hat{\mathcal{R}}_{\hat{z}}(\phi_s) \, \hat{\mathcal{R}}_{\hat{x}} \! \left(\theta(y - \bar{y})\right) \begin{pmatrix} \cos(Qx + \phi_t) \\ \sin(Qx + \phi_t) \\ 0 \end{pmatrix},
\end{equation}
where $\hat{\mathcal{R}}_{\hat{e}}(\alpha)$ is the rotation matrix around $\hat{e}$ by the angle $\alpha$. For $y \ll \bar{y}$ and $y \gg \bar{y}$, \eqref{eq:Ansatz} reduces to the positive and negative chirality spirals \eqref{eq:Cycloids} with $\phi \equiv \phi_+ = \phi_t +\phi_s$ and $\phi \equiv \phi_- = \phi_t - \phi_s$. The function $\theta(y - \bar{y})$ is determined by minimizing the last two terms of the Hamiltonian \eqref{eq:H0} (see SI for details), resulting in
\begin{equation}
\label{eq:Theta}
\hat{n}_z = \cos\theta(y - \bar{y}) = - \tanh\! \left( \frac{y-\bar{y}}{\lambda_0} \right) ,
\end{equation}
where $\lambda_0 = \sqrt{\lvert J_\perp \rvert / 2 K_z}$ is the DW width. Therefore, a meron DW is simply a Bloch/Néel wall in $\hat{n}$ (see Fig.~\ref{fig:StaticDW}(a), right inset). The surface tension $\sigma = \sqrt{2 K_z \lvert J_\perp \rvert}$ also resembles that of a DW in a ferromagnet with spin stiffness $J_\perp$ and easy axis anisotropy $K_z$.

The domain wall described above consists of a periodic array of meron strings extending along $z$ and spaced by $\Delta x=\pi/Q$, in agreement with \cite{Li12}. Figure~\ref{fig:StaticDW} shows the magnetization texture in the $xy$-plane, which repeats for every fixed $z$. Inside the closed gray contour, ${\bf m}$ wraps the lower hemisphere, giving rise to a meron with topological charge $\mathcal{Q} = -1/2$. Such a meron has vorticity $n = 1$ and helicity angle $\phi_s = 0$. In fact, circulating counterclockwise around the meron string, ${\bf m}$ rotates by $2\pi$, and its projection on the $xy$-plane points along the radial direction close to the center. In general, the merons within the DW have the same vorticity $n = 1$ but alternating topological charge $\mathcal{Q} = \pm 1/2$ and helicity angle $\varphi_j = \phi_s+\pi j$. If we reverse the domain chiralities, the meron vorticity becomes $n = -1$. We note that changes in $\phi_s$, $\phi_t$ and $\bar{y}$ correspond to rotations of ${\bf m}$ in the $xy$-plane, translations in the $x$-direction and translations in the $y$-direction, respectively. Thus, $\phi_s$, $\phi_t$ and $\bar{y}$ are the collective coordinates associated with the continuous symmetries of the Hamiltonian \eqref{eq:H0}.

Our analysis concerns the DWs whose plane contains the spiral wave vector ${\bf Q}$.
A wall with normal $\hat{e}_\perp$ forming an angle $\delta$ with ${\bf Q}$ presents two meron strings per spiral period spaced by $\pi/(Q\sin\delta)$, similar to vortex DWs \cite{Li12}. Hence, only Hubert DWs (i.e. $\delta = 0$) are meron-free, and we expect meron DW motion for $\delta = \pi/2$ to capture the essence of the dynamics for all other wall orientations.

\paragraph{Electric field-driven dynamics. ---}
Now we consider a DW in the presence of a uniform electric field ${\bf E} = E_y \hat{y}$. Inverse Dzyaloshinskii-Moriya effect that couples ${\bf E}$ and ${\bf m}$ is described by the Hamiltonian density \cite{Mostovoy06,Katsura05,Sergienko06}:
\begin{equation}
\label{eq:H'}
\mathcal{H}_\mathrm{DM} = -\gamma {\bf E} \cdot \left[ ({\bf m} \cdot \nabla) {\bf m} - {\bf m} (\nabla \cdot {\bf m}) \right] .
\end{equation}
Spiral domains with opposite chirality $\chi$ no longer have the same energy density because $\mathcal{H}_\mathrm{DM} = -\chi \gamma E_y Q$, and the favorable domain grows at the expense of the other, that is the meron DW moves in the $y$-direction.

The dynamics of magnetization is described by the following Lagrangian density and Rayleigh dissipation functional density \cite{Landau35, Gilbert04, Duine10}
\begin{equation}
\label{eq:LeR}
\mathcal{L} = {\bf A}({\bf m}) \cdot \dot{{\bf m}} - \mathcal{H}_\mathrm{tot}, \qquad \mathcal{R} = \frac{\alpha}{2} \dot{{\bf m}}^2 ,
\end{equation}
where ${\bf A}({\bf m})$ satisfies $\nabla_{\bf m} \times {\bf A}({\bf m}) = {\bf m}$, $\mathcal{H}_\mathrm{tot} = \mathcal{H} + \mathcal{H}_\mathrm{DM}$, $\alpha \ll 1$ is the Gilbert damping constant and the time $t$ is redefined according to $t' = t/S$ with $S$ denoting the spin. The dynamics described by \eqref{eq:LeR} can be solved exactly only in the simplest cases \cite{Schryer74}. Therefore, we apply the collective coordinates approach \cite{Tretiakov08, Clarke08}. In such a framework, the dynamics is formulated in terms of collective coordinates $\xi_I(t)$, so that ${\bf m}(t,{\bf r}) = {\bf m}(\{\xi_I(t)\},{\bf r})$. Although ${\bf m}(t,{\bf r})$ has an infinite number of $\xi_I(t)$, only the \textit{soft} modes with long relaxation times compared to the characteristic time of the dynamics are relevant. The other \textit{hard} modes adjust adiabatically to their equilibrium values. In particular, the low-field (i.e. long time) dynamics is dominated by the \textit{zero} modes corresponding to the continuous symmetries of the Hamiltonian \eqref{eq:H0}, that is $\phi_s$, $\phi_t$ and $\bar{y}$.

From \eqref{eq:LeR}, descend the equations of motion for the collective coordinates \cite{Tretiakov08, Clarke08}
\begin{equation}
\label{eq:Thiele}
\alpha \Gamma_{IJ} \dot{\xi}_J - G_{IJ} \dot{\xi}_J = F_I ,
\end{equation}
where $I$ and $J$ run over the soft modes, and the damping matrix $\Gamma_{IJ} = \Gamma_{JI}$, the gyrotropic matrix $G_{IJ} = - G_{JI}$ and the conservative forces $F_I$ acting on $\xi_I$ descend from $\mathcal{R}$, ${\bf A}({\bf m}) \cdot \dot{{\bf m}}$ and $\mathcal{H}_\mathrm{tot}$, respectively (see SI for details).

The conservative forces due to the electric field $E_y$ are $F_{\bar{y}} = q E_y L_x L_z$, $F_{\phi_s} = F_{\phi_t}=0$, where $q = 2 Q \gamma$ is the electric charge density of the DW and the dimensions of the system are $L_x\times L_y\times L_z$. In the simplest case, where the wall contains an even number of merons with alternating topological charges $\mathcal{Q}_i$, $\bar{y}$ does not interact with $\phi_s$ and $\phi_t$. Indeed, $G_{\bar{y} \phi_t} = 4 \pi L_z \mathcal{Q}_\mathrm{tot} / Q$ is proportional to the total topological charge of the DW $\mathcal{Q}_\mathrm{tot}=\sum_i \mathcal{Q}_i = 0$. (The general case $\mathcal{Q}_\mathrm{tot} \neq 0$ is discussed below.) The other components of $\Gamma_{IJ}$ and $G_{IJ}$ that couple $\bar{y}$ with $\phi_s$ and $\phi_t$ vanish for the mirror symmetries $m'_{x,y}$ of the merons (see Fig.~\ref{fig:StaticDW}(a)). Therefore, $\phi_s$ and $\phi_t$ have a trivial dynamics, and, for the zero modes, \eqref{eq:Thiele} reduces to $\alpha \Gamma_{\bar{y}\bar{y}} \dot{\bar{y}} = F_{\bar{y}}$. All the previous results do not rely on the particular form of the DW ansatz. Thus, they are expected to hold beyond $K_z \ll J_x$. Using \eqref{eq:Ansatz} and \eqref{eq:Theta}, the damping constant of $\bar{y}$ reads $\alpha \Gamma_{\bar{y}\bar{y}} = \beta L_x L_z$, where $\beta = \alpha / \lambda_0$, and the low-field DW velocity takes the form
\begin{equation}
\label{eq:Steady}
\dot{\bar{y}} = Q \sqrt{\frac{2 \lvert J_\perp \rvert}{K_z}} \frac{\gamma E_y}{\alpha} \equiv v_{\infty} .
\end{equation}
At low electric fields, the long-time dynamics consists of the rigid translational motion of the DW with a constant velocity determined by the balance between the electric field and the dissipation. The system reaches this steady motion regime after a transient in which the hard modes adjust to their new equilibrium values. Nevertheless, at the first order in $E_y$, the steady-state velocity \eqref{eq:Steady} is not affected by the presence of hard modes (see SI).

\begin{figure}[t]
\includegraphics[width=1\linewidth]{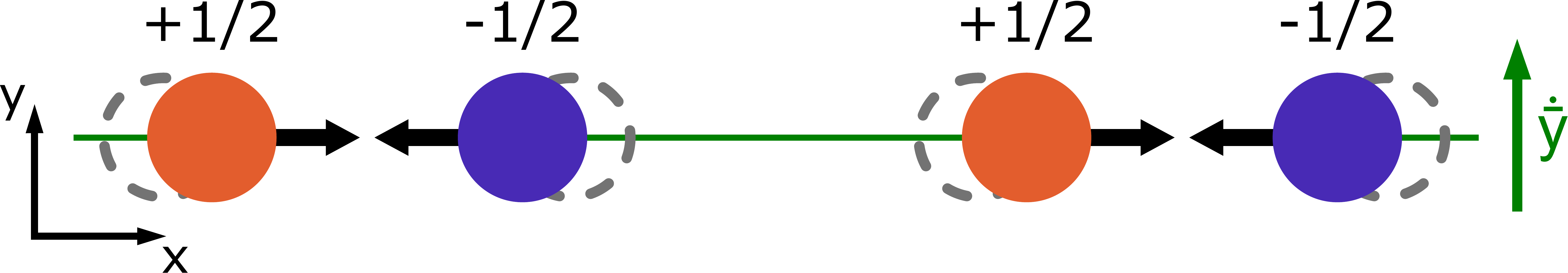}
\caption{\label{fig:Dimer}Meron DW moving with velocity $\dot{\bar{y}}$. The Lorentz-like gyrotropic force pushes oppositely charged merons in opposite directions, leading to dimerization.
}\end{figure}

To describe the transient dynamics, we need to include the momentum conjugate to $\bar{y}$, which it is coupled to via the gyrotropic matrix $G_{IJ}$. For the $i$-th meron, $G_{IJ}$ couples its translational modes along the $x$ and $y$ directions. Reparameterizing these modes in terms of the meron center $(\bar{x}, \bar{y})$, we get $G_{\bar{x} \bar{y}}^i = 4 \pi L_z \mathcal{Q}_i$. Thus, considering each term in \eqref{eq:Thiele} as a force, the meron experiences a gyrotropic force ${\bf F}_G^i =  4 \pi L_z \mathcal{Q}_i \, [(\dot{\bar{x}}, \dot{\bar{y}}, 0) \times \hat{z}]$ analogous to the Lorentz force \cite{Tretiakov08}. In particular, as the wall moves with a velocity $\dot{\bar{y}} > 0$, merons with positive $\mathcal{Q}_i$ are pushed to the right, while those with negative $\mathcal{Q}_i$ are pushed to the left, until ${\bf F}_G^i$ is balanced by the restoring force from $J_x$ in \eqref{eq:H0} (see Fig.~\ref{fig:Dimer}). Hence, the meron DW dimerizes when it moves. Since this effect is due to the gyrotropic force, we expect the dimerization mode $d$ to have $G_{d\bar{y}} \neq 0$ and enter the effective Lagrangian as the momentum conjugate to $\bar{y}$.

The alternating translations of merons along the $x$ direction representing dimerization can be incorporated in \eqref{eq:Ansatz} via $\phi_t \rightarrow \phi_t + f(t, x, y - \bar{y}(t))$, where $f$ is determined by solving the variational problem of \eqref{eq:LeR} (see SI). In the steady motion regime, $f$ has no explicit time dependence and $\bar{y}(t) = v_{\infty}t$. At the first order in $v_{\infty}$, we obtain
\begin{equation}
\label{eq:f}
f(x, y - v_{\infty}t) = d_{\infty} \frac{\sin(Qx + \phi_t)}{\cosh({(y - v_{\infty}t)/\lambda_0})} ,
\end{equation}
where $d_{\infty} = m v_{\infty}$ determines dimerization in the steady motion regime and $m^{-1} = 5 J_x \lambda_0 Q^4$.
Replacing $d_{\infty}$ with $d(t)$ in \eqref{eq:f} defines $f(t, x, y - \bar{y}(t))$ and therefore the dimerized configuration at time $t$ (see SI for an alternative definition). We observe that dimerization is associated with a net magnetic moment within the DW
\begin{equation}
\label{eq:NetMom}
\int{dV {\bf m}} = - \frac{\pi}{2} \lambda_0 L_x L_z d \left(\cos\phi_s, \sin\phi_s, 0\right),
\end{equation}
which is the momentum conjugate to the rotations of ${\bf m}$ around $\hat{e} = \left(\cos\phi_s, \sin\phi_s, 0\right)$. Such a rotation within the DW \eqref{eq:Ansatz} corresponds to its motion in the $y$-direction.

At the first order in $E_y$, \eqref{eq:Thiele} takes the form
\begin{equation}
\label{eq:EqYD1}
\begin{pmatrix} \beta & 0 \\ 0 & \beta \lambda_0^2 \end{pmatrix} \begin{pmatrix} \dot{\bar{y}} \\ \dot{d} \end{pmatrix} - \begin{pmatrix} 0 & -1 \\ 1 & 0 \end{pmatrix}  \begin{pmatrix} \dot{\bar{y}} \\ \dot{d} \end{pmatrix} = \begin{pmatrix} q E_y \\ - d / m \end{pmatrix},
\end{equation}
This equation descends from the effective Lagrangian and Rayleigh dissipation function
\begin{equation}
\label{eq:EffLeR}
L = \dot{\bar{y}} d + q E_y \bar{y} - \frac{d^2}{2 m} , \qquad R = \frac{1}{2} \beta \dot{\bar{y}}^2 + \frac{1}{2} \beta \lambda_0^2 \dot{d}^2 .
\end{equation}
We observe that $d$ is the  momentum conjugate to $\bar{y}$ (i.e. $d = \partial L / \partial \dot{\bar{y}}$), as anticipated.

Neglecting second-order terms in $\alpha \ll 1$ and assuming $\dot{\bar{y}}/v_{\infty} \gg \alpha^2$, the equations of motion \eqref{eq:EqYD1} become
\begin{equation}
\label{eq:EqYD2}
\dot{d} = m \ddot{\bar{y}} = q E_y - \beta \dot{\bar{y}} , \qquad d = m \dot{\bar{y}}.
\end{equation}
At low electric fields, the meron DW behaves as a particle with mass $m$, charge $q$ and viscous friction coefficient $\beta$. The dimerization mode plays the role of the momentum.

Solving \eqref{eq:EqYD2} with the initial condition $\dot{\bar{y}} = v_0$ we get
\begin{equation}
\label{eq:Transient}
\dot{\bar{y}} = v_0 e^{-t/\tau} + v_{\infty} \left(1 - e^{-t/\tau}\right),
\end{equation}
where $\tau = m/\beta$
denotes the relaxation time of the dimerization mode $d(t)$, and $v_\infty$ is the steady-state velocity \eqref{eq:Steady}. Such a result captures both the transient and the steady motion regime. However, it cannot resolve the dynamics at timescales $t \ll \tau$ (e.g. the response to a pulse) because other modes may be active. In addition, at higher orders in $E_y$, DW deformations associated with hard modes may affect the dynamics. In particular, the effective mass and damping coefficient depend on the DW width $\lambda$. Hence, the high-field dynamics is sensitive to the DW breathing mode that modulates $\lambda$.

Thus, we treat $\lambda$ as one of the collective coordinates by replacing $\lambda_0$ with $\lambda(t)$ in \eqref{eq:Theta} and \eqref{eq:f}. Since \eqref{eq:f} does not rely on the particular form of $v_\infty$, it continues to capture the dynamics up to the second order in the dimerization for $K_z \ll J_x$. Neglecting second-order terms in $\alpha$, $\dot{d}$ and $\dot{\lambda}$, the equations of motion for $\bar{y}$, $d$ and $\lambda$ take the form (see SI for details)
\begin{equation}
\label{eq:EqRel}
m \frac{d}{dt}(\gamma_c \dot{\bar{y}}) = q E_y - \beta \gamma_c \dot{\bar{y}} , \quad d = m \gamma_c \dot{\bar{y}} , \quad \lambda = \frac{\lambda_0}{\gamma_c},
\end{equation}
where $\gamma_c^{-1} = \sqrt{1 - \dot{\bar{y}}^2/c_\mathrm{m}^2}$, and $c_\mathrm{m} = \sqrt{5 \lvert J_\perp \rvert J_x Q^4}$ is the DW limiting velocity. Remarkably, it coincides with the maximum magnon speed \cite{Zvezdin79, Caretta20, Du16}. As the wall moves, its width is contracted by the Lorentz-like factor $\gamma_c$. Since $m$ and $\beta$ are inversely proportional to $\lambda$, they are multiplied by $\gamma_c$, leading to a ``relativistic'' DW dynamics, where $c_\mathrm{m}$ plays the role of the speed of light. Indeed, \eqref{eq:EqYD2} goes into \eqref{eq:EqRel} through the substitution $\dot{\bar{y}} \rightarrow \gamma_c \dot{\bar{y}}$. The DW steady-state velocity \eqref{eq:Steady} is modified by the contraction of $\lambda$:
\begin{equation}
\label{eq:SteadyRel}
v_{\infty}^{\mathrm{rel}} = \frac{v_\infty}{\sqrt{1 + v_\infty^2 / c_\mathrm{m}^2}} < c_\mathrm{m}.
\end{equation}
It approaches $v_\infty$, \eqref{eq:Steady}, at low fields (i.e. when $v_\infty \ll c_\mathrm{m}$) and saturates toward the magnon speed $c_\mathrm{m}$ at high fields.

We corroborate the results with atomistic spin dynamics simulations \cite{Landau35, Gilbert04, Skubic08}. First, we simulate a $J_1$-$J_2$ model with competing nearest and next-nearest neighbor interactions along $x$ and ferromagnetic interactions in the $yz$-planes (see SI, Fig.~S1, Fig.~S2 and Movie~1 for details). Remarkably, \eqref{eq:Steady} and \eqref{eq:SteadyRel} also hold when the interactions along $y$ or $z$ are antiferromagnetic, although the expression for the magnon speed differs. This case, relevant to many real multiferroic perovskites \cite{Mochizuki09a, Mochizuki09b, Mochizuki10a, Mochizuki10b, Mochizuki10c, Mochizuki11a, Mochizuki11b, Matsubara15}, is discussed in the final section. The structure and electric field-driven dynamics of the minimum energy meron DW (see Fig.~\ref{fig:StaticDW}) are in excellent agreement with simulations.

\paragraph{Defects within the meron wall. ---}
Equation~\eqref{eq:Theta} admits two solutions $\pm\theta(y - \bar{y})$ where the spin rotation axis $\hat{n}$ rotates in opposite directions. If solutions $\pm\theta$ are realized across a boundary $x = x^*$ lying between two merons (dashed line in Fig.~\ref{fig:Defect}), that leads to a low-energy defect within the DW, where two neighboring merons have the same topological charges $\mathcal{Q}_i=\mathcal{Q}_{i+1}$. This gives rise to a meron DW with non-alternating $\mathcal{Q}_i$ (e.g. with signature $\{...-+--+-\,...\}$, Fig.~\ref{fig:Defect}), in contrast to the previously considered minimum energy meron DW with alternating $\mathcal{Q}_i$ (i.e. $\{...+-+-\,...\}$, Fig.~\ref{fig:StaticDW}).

\begin{figure}[b]
\centering
\includegraphics[width=1\linewidth]{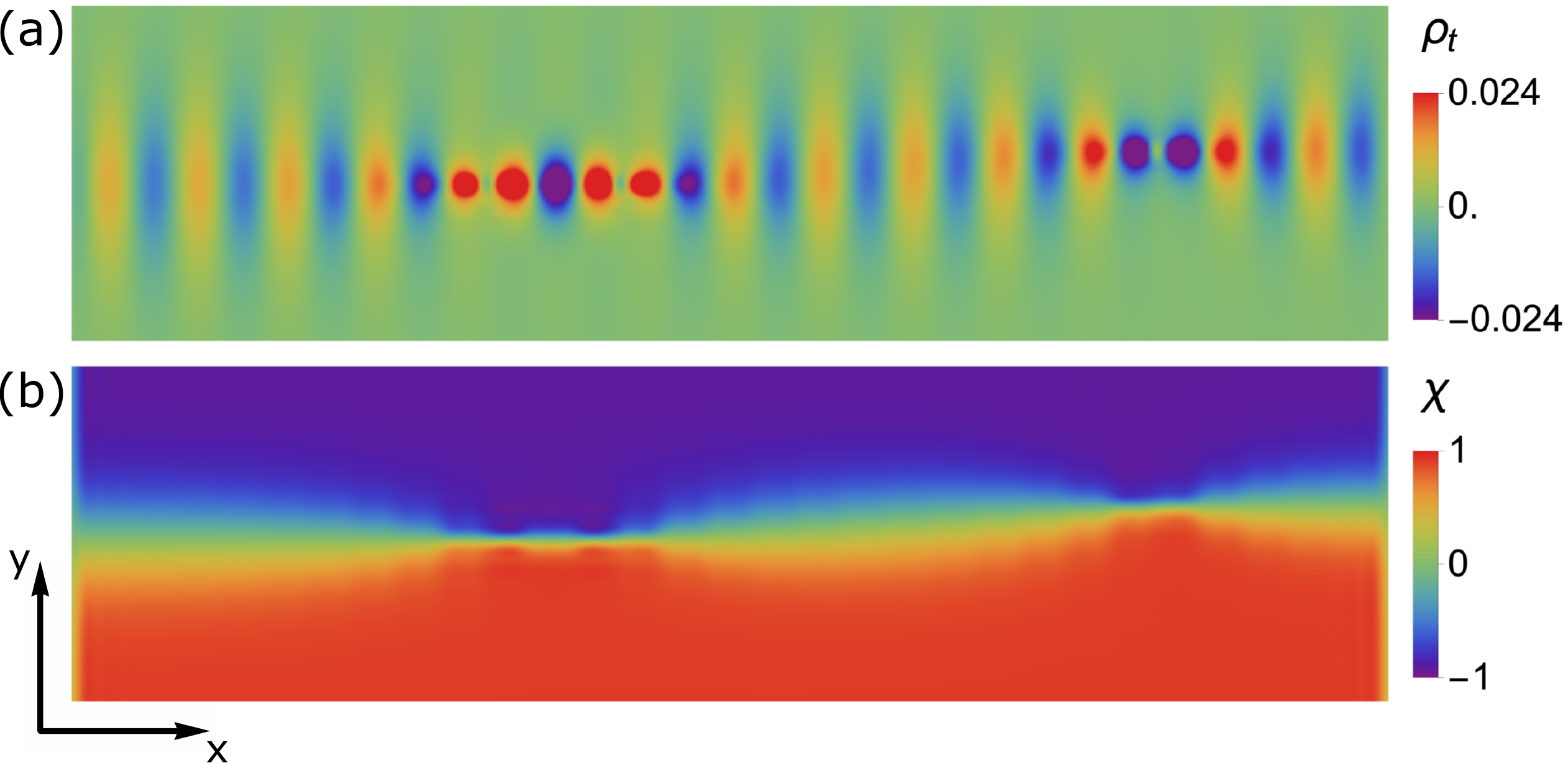}
\caption{\label{fig:Domains}Meron DW obtained by relaxing a paramagnetic configuration (see Fig.~S3 for the entire domain structure, -- Fig.~\ref{fig:StaticDW} and Fig.~\ref{fig:Defect} for the internal DW structure). Panels (a) and (b) show the topological charge density $\rho_\mathrm{t} = {\bf m} \cdot \left[\partial_x {\bf m} \times \partial_y {\bf m}\right]/4\pi$ and the generalized chirality $\chi \equiv P_y/\gamma Q = \hat{z} \cdot \left[{\bf m} \times \partial_x {\bf m}\right]/Q$. Defects in the alternating order of meron topological charges correspond to segments where the wall is sharp. The DW has a net topological charge $\mathcal{Q}_\mathrm{tot}$, as the numbers of positive and negative defects differ.
}\end{figure}

By relaxing a 2D paramagnetic state via the LLG dynamics \cite{Landau35, Gilbert04, Skubic08} (see SI for details), we obtain several spiral domains \eqref{eq:Cycloids}, separated by meron DWs (see Fig.~\ref{fig:Domains} and Fig.~S3). Meron topological charges $\mathcal{Q}_i$ tend to alternate, but defects in this alternating order are also present (see Fig.~\ref{fig:Domains}(a)). Consequently, we expect these defects to exist in real systems, and studying how they affect the dynamics is an essential problem.

Near a defect, the energy increases because the dominant term of the Hamiltonian density \eqref{eq:H0}, proportional to $J_x$, is not minimized (i.e., for a fixed $y$, the configuration is not a single spiral with wave vector $\pm Q$, see Fig.~\ref{fig:Defect}(b)). Consequently, merons shrink in the $y$-direction to reduce the energy, decreasing the DW width to $\lambda \sim \sqrt{\lvert J_\perp \rvert/J_x Q^4}$ below the ideal DW width $\lambda_0 = \sqrt{\lvert J_\perp \rvert/2 K_z}$. Hence, the effective viscous friction coefficient $\beta = \alpha/\lambda$ is enhanced near defects, which fall behind as the meron DW moves (see SI, Fig.~S4). On average, $\beta$ is enhanced by the factor $b = (1/L_x) \int_0^{L_x}{dx \:  \lambda_0/\lambda(x)}$.

When the numbers of positive and negative defects are different, the DW has a net topological charge $\mathcal{Q}_\mathrm{tot}$, and $G_{\bar{y} \phi_t} = 4 \pi L_z \mathcal{Q}_\mathrm{tot} / Q \neq 0$. As the wall moves, $\phi_t$ experiences a force $-G_{\bar{y} \phi_t} \dot{\bar{y}}$ and $\phi_s$ -- a force $- \alpha \Gamma_{\phi_s \phi_t } \dot{\phi}_t$. Therefore, $\mathcal{Q}_\mathrm{tot}$ couples the DW position $\bar{y}$ with the phasons of the positive and negative chirality domains $\phi_\pm = \phi_t \pm \phi_s$. The magnetization rotates in the entire system unless $\phi_t$ and $\phi_s$ are pinned (e.g. by surface effects, anisotropy or disorder). Solving the equations of motion \eqref{eq:Thiele} for $\phi_s$, $\phi_t$ and $\bar{y}$ while neglecting pinning effects, we get
\begin{eqnarray}
\dot{\bar{y}} &=& v_\infty \left(b + \frac{1}{\alpha^2} \frac{L_y \lambda_0}{\bar{y}(L_y - \bar{y})} \left(\frac{2 \mathcal{Q}_\mathrm{tot}}{N}\right)^2 \right)^{-1}, \nonumber\\
\dot{\phi}_+ &\equiv& \dot{\phi}_t + \dot{\phi}_s = -\frac{2 \mathcal{Q}_\mathrm{tot}}{N} \frac{\dot{\bar{y}}}{\alpha \bar{y}}, \nonumber\\
\dot{\phi}_- &\equiv& \dot{\phi}_t - \dot{\phi}_s = -\frac{2 \mathcal{Q}_\mathrm{tot}}{N} \frac{\dot{\bar{y}}}{\alpha (L_y - \bar{y})}, \label{eq:SolDef}
\end{eqnarray}
where $N = L_x Q / \pi$ is the total number of merons within the DW, $\dot{\phi}_+$ and $\dot{\phi}_-$ are the precession speeds of ${\bf m}$ in the positive and negative chirality domains \eqref{eq:Cycloids}, respectively. As the wall moves, ${\bf m}$ rotates in each spiral domain, reducing the DW mobility by a factor that depends on its position $\bar{y}$ and total topological charge $\mathcal{Q}_\mathrm{tot}$. Such a non-local dynamics resemble that studied in Hubert DWs \cite{Foggetti22}, which have $\mathcal{Q}_\mathrm{tot}=0$, but the origin is quite different. In fact, the meron DW position $\bar{y}$ and the spiral phases $\phi_\pm$ are three different zero modes coupled together (a Hubert DW instead has only two zero modes $\phi_\pm$). In particular, a meron DW can move even when $\phi_\pm$ are pinned, and ${\bf m}$ do not rotate in the entire system.
Also, unlike a Hubert DW, the velocity of a meron DW with $\mathcal{Q}_\mathrm{tot} \neq 0$ reaches its maximum at the system center and goes to zero as it approaches the surfaces $y = 0$ and $y = L_y$.  We note that the result \eqref{eq:SolDef} holds in the bulk (i.e. $\lambda_0 \ll \bar{y} \ll L_y - \lambda_0$), and surface effects are expected to prevent the wall from getting trapped inside the system (see SI, Fig.~S5).
Due to non-local dynamics, two distant DWs with net topological charges can affect each other’s motion (see SI for details). Hence, the DW velocities depend on the entire domain structure, similar to Hubert DWs. It would be interesting to analyze the dynamics of DWs in noncentrosymmetric helical magnets, which present analogues of both Hubert and meron walls \cite{Schoenherr18}.

In the dynamics described by \eqref{eq:SolDef}, merons translate in the $x$-direction with a velocity $\dot{\bar{x}} = -\dot{\phi}_t/Q \neq 0$. Therefore, merons are destroyed at one end of the wall, and new ones are created at the other (see Movies~2--4). Since the energy is higher when the boundary is cutting through a meron and lower when it is between two merons, such a process may affect the dynamics, exerting a force on $\phi_t$ and possibly exciting new modes. We study the effects of meron creation and destruction by performing numerical simulations (see SI, Fig.~S5 and Movies~2--4).  With open boundary conditions, the DW velocity oscillates around \eqref{eq:SolDef} with an amplitude proportional to the energy of created and destroyed merons. When a defect is destroyed, $\dot{\bar{y}}$ drops sharply, and $\mathcal{Q}_\mathrm{tot}$ changes because merons are created with alternating $\mathcal{Q}_i$. Except for these events, which occur on a short timescale compared to the characteristic time of the dynamics, \eqref{eq:SolDef} captures the averaged $\dot{\bar{y}}$ (see SI, Fig.~S5). Consistently, ruling out meron creation and destruction by imposing periodic boundary conditions in the $x$-direction, simulations are in very good agreement with \eqref{eq:SolDef} for $\lambda_0 \ll \bar{y} \ll L_y - \lambda_0$.

\paragraph{Topological charge control. ---}
An external magnetic field $H_z \hat{z}$ profoundly affects the dynamics of meron DWs. Indeed, $H_z$ induces a conical component in the spirals \eqref{eq:Cycloids}, which becomes ${\bf m}_\chi^{\vartheta_c}(x) = \sin\vartheta_c \, {\bf m}_\chi(x) + \cos\vartheta_c \, \hat{z}$, where $\vartheta_c \in (0,\pi)$ denotes the conical angle and $\vartheta_c = \pi/2$ gives a flat spiral \eqref{eq:Cycloids} (see Fig.~\ref{fig:TopControl}(a)). Minimizing the energy \eqref{eq:H0} with the Zeeman term $\mathcal{H}_\mathrm{Z} = - \mu H_z  m_z$, where $\mu$ denotes the magnetic moment of a single spin, we get
\begin{equation}
\label{eq:ConicalTheta}
\cos\vartheta_c = \frac{\mu H_z}{2 K_z + J_x Q^4}.
\end{equation}
Within a positive meron, the polar angle $\vartheta$ varies between $0$ and $\vartheta_c$, while within a negative meron, $\vartheta \in (\vartheta_c,\pi)$ (see Fig.~\ref{fig:TopControl}(b)). Thus, the topological charges become
\begin{equation}
\label{eq:TopQ}
\mathcal{Q}_+ = \frac{1 - \cos\vartheta_c}{2}, \qquad \mathcal{Q}_- = -\frac{1 + \cos\vartheta_c}{2},
\end{equation}
respectively, where $\cos\vartheta_c$ is given by \eqref{eq:ConicalTheta}. Therefore, $H_z$ induces a topological charge on the wall so that even the minimum energy DW, with alternating merons, has a net topological charge and exhibits non-local dynamics.
This is illustrated in Movie 5, where $H_z$ is turned on after the wall reaches the steady motion regime, leading to a sharp deceleration due to spin precession in the entire system.

\paragraph{Helicity control. ---}
As the spin rotation axis $\hat{n}$ rotates across the meron DW, the ferroelectric polarization ${\bf P} \propto [\hat{n}\times{\bf Q}]$ rotates with it, endowing the wall with a net $P_z$ (unless $\hat{n}$ passes through the $\bf{Q}$ direction).
Such a $P_z$ reverses across defects, which can facilitate their experimental detection, e.g. using second harmonic generation.

Within a meron DW, away from defects, the rotation of $\bf P$ induces an average $P_z$:
\begin{equation}
\label{eq:Pol}
P_z = -\frac{1}{V}\int{dV \: \frac{\partial \mathcal{H}_\mathrm{DM}}{\partial E_z}} = \frac{k}{L_y} \cos\phi_s,
\end{equation}
where $k = \pi\gamma(\lambda Q + 1/2)$. Hence, the meron helicity angle $\phi_s$ determines $P_z$ and is sensitive to an external electric field $E_z$. The force exerted by $E_z$ on the minimum energy DW takes the form $F_{\phi_s} = - k E_z L_x L_z \sin\phi_s$. When $\phi_t$ is pinned (e.g. by surface effects) or does not interact with $\phi_s$ (i.e. the wall is at the center of the system $\bar{y} = L_y/2$), $\phi_s$ satisfies $\dot{\phi}_s =  -(1/\tau_s)\sin\phi_s$, where $\tau_s = \alpha L_y / k E_z$ has the same sign as $E_z$. The solution
\begin{equation}
\label{eq:PolSol}
P_z = \frac{k}{L_y} \tanh{\frac{t}{\tau_s}}
\end{equation}
describes the switching of $P_z$ between $+k/L_y$ and $- k/L_y$ ($\phi_s$ between $0$ and $\pm\pi$) in an infinite time. Therefore, we expect the switching time of $P_z$ to be strongly influenced by fluctuations (e.g. thermal fluctuations).

Taking into account that $P_z$ is inverted across a defect (i.e. $\phi_s$ changes by $\pm\pi$), the result can be generalized to a meron DW with defects. Now, if $\bar{y} \neq L_y/2$ and $\mathcal{Q}_\mathrm{tot} \neq 0$, the DW moves under the action of $E_z$. In fact, the field acts indirectly on $\bar{y}$ because the zero modes interact with each other.

\paragraph{Defects within meron strings. ---}
Up to this point, we have considered quasi-2D magnetization textures, where the meron DWs consist of an array of meron strings extending along $z$. Now, we study point defects within the meron strings. These are similar to the defects discussed above, but the solutions $\pm\theta$ are realized across a boundary $z = z^*$ instead of $x = x^*$ (see Fig.~\ref{fig:TopControl}(c,$\,$f)). Hence, the topological charge of each meron string is reversed across $z = z^*$, resulting in an array of point defects, one for each string. Such defects are hedgehogs with alternating magnetic monopole charges because ${\bf m}$ wraps the upper and the lower hemisphere on opposite sides of $z = z^*$, and the meron helicities alternate within the wall. Meron strings do not carry net magnetic monopole charges since hedgehogs are compensated by charges arising at the intersections between strings and system surfaces. By relaxing a paramagnetic state, such arrays of hedgehogs arise naturally. A generic defect line within the DW plane presents two hedgehogs per spiral period. Thus, a single hedgehog corresponds to a defect line that extends along the spiral wave vector by $\pi/Q$ (see Fig.~\ref{fig:TopControl}(g)).
The lengthscale of a hedgehog is set by the lattice constant. Hence, point defects are strongly pinned to the lattice by Peierls-Nabarro barriers \cite{Peierls40, Nabarro47} and give rise to intrinsic DW pinning.

\paragraph{Antiferromagnetic case. ---}
Thus far, we have considered materials with a ferromagnetic $J_{\perp}$ (e.g. single-layer NiI$_2$, NiBr$_2$, and VI$_2$ \cite{Song22, Sødequist23}). If $J_\perp$ is antiferromagnetic at least along one direction, as in TbMnO$_3$, MnWO$_4$, and CuO \cite{Kimura03, Tokura10}, the spin texture is described by the Néel vector ${\bf n}({\bf r}) = ({\bf m}_1({\bf r}) - {\bf m}_2({\bf r}))/2$ and the net magnetization ${\bf m}({\bf r}) = {\bf m}_1({\bf r}) + {\bf m}_2({\bf r})$, where ${\bf m}_1({\bf r})$ and ${\bf m}_2({\bf r})$ are the magnetization fields in the two antiferromagnetic sublattices. Integrating out ${\bf m}({\bf r})$, the Lagrangian density and Rayleigh dissipation functional density \eqref{eq:LeR} become \cite{Dasgupta21}
\begin{equation}
\label{eq:LeRAf}
\mathcal{L} = \frac{\dot{{\bf n}}^2}{8 J_\perp N_\mathrm{AF}} - \mathcal{H}_{\bf n}, \qquad \mathcal{R} = \frac{\alpha}{2} \dot{{\bf n}}^2 ,
\end{equation}
where $J_\perp$ is antiferromagnetic in $N_\mathrm{AF} = 1,2,3$ directions, and the Hamiltonian density $\mathcal{H}_{\bf n}$ is identical to the case of ferromagnetic $J_{\perp}$ (see \eqref{eq:H0} and \eqref{eq:H'}), with the replacement ${\bf m} \rightarrow {\bf n}$. 
Therefore, a minimum energy meron DW is still described by \eqref{eq:Ansatz}, which is now an expression for ${\bf n}(x,y)$, and low-energy defects are still present (see Fig.~\ref{fig:TopControl}(d--g)). However, topological charges mutually cancel in the two antiferromagnetic sublattices, and there is no gyrotropic force. Hence, the meron DW no longer dimerizes, and its motion cannot induce spin precession in the entire spiral domains, even in the presence of defects.
The equations of motion for $\bar{y}$ and $\lambda$ still take the form \eqref{eq:EqRel}, but with the replacement $5 J_x Q^4 \rightarrow 4 J_\perp N_\mathrm{AF}$ that affects the effective mass $m$ and the limiting velocity $c_\mathrm{m}$ of the DW. Defects still enhance the viscous friction coefficient $\beta$ and lead to DW pinning due to the lattice potential (Peierls-Nabarro barriers), thus significantly contributing to coercivity.

We simulate meron DW motion in TbMnO$_3$ using LLG equations \cite{Landau35, Gilbert04} and the microscopic spin Hamiltonian proposed by Mochizuki et al. \cite{Mochizuki09a, Mochizuki09b, Mochizuki10a, Mochizuki10b, Mochizuki10c, Mochizuki11a, Mochizuki11b, Matsubara15} (see SI for details). The steady-state velocity of the wall is well described by \eqref{eq:Steady} and \eqref{eq:SteadyRel}, with the replacement introduced above. In particular, the low-field DW mobility is $\mu_\mathrm{DW} = v_\infty/E = 1 \cdot 10^{-5} \,\mathrm{m^2 \, s^{-1} V^{-1}}$ in the absence of defects. In MnWO$_4$, the ferroelectric polarization is weaker, and we expect the DW mobility to be an order of magnitude lower. Defects within the meron DWs arise naturally in our simulations of TbMnO$_3$ when starting from a random configuration. The depinning field $E_\mathrm{pin}$ of a meron DW depends on the concentration of defects and sets the coercive field for the polarization switching. Simulating a configuration where the defect lines intersect the meron strings (see Fig.~\ref{fig:TopControl}(f)) and have an average distance $\Delta z$ of 46 lattice constants, we obtain the coercive field $E_\mathrm{pin} \approx 10^7 \,\mathrm{Vm^{-1}}$. Additional simulations confirm that $E_\mathrm{pin}$ is directly proportional to the concentration of defects. Consequently, the presence of a defect line approximately every 460 lattice constants corresponds to a zero-temperature coercive field, $E_\mathrm{pin} \approx 10^6 \,\mathrm{Vm^{-1}}$, comparable to the fields commonly applied in experiments.

\paragraph{Conclusions. ---}
We have shown that centrosymmetric spiral magnets exhibit DWs consisting of an array of meron strings, which result from the twisting of the spiral plane across the wall. Such a twist is concomitant with a rotation of the ferroelectric polarization by 180$^\circ$. Hence, meron DWs are also ferroelectric walls and can be driven by an electric field. The resulting dynamics is intimately connected with the magnetization texture topology and depends on whether the magnetization or the Néel vector rotates within the spiral. Still, the meron DW structure is the same in both cases.

Meron topological charges alternate along a minimum energy wall. Nevertheless, low-energy defects in this alternating charge order appear during domain nucleation and can lead to meron DWs with net topological charge. Another type of defect separates the segments of a meron string with opposite topological charges. This hedgehog point defect, analogous to a Bloch point, carries magnetic monopole charge and pins the wall to the lattice. In both cases, the ferroelectric polarization $P_z$ reverses across the defect. Consequently, polarization force microscopy and second harmonic generation experiments could detect it. The results also apply to single-layer spiral multiferroics (e.g. NiI$_2$ \cite{Song22}), except for hedgehog defects that are only possible in 3D.

If the net topological charge of the meron DW is zero, its motion resembles that of a ``relativistic'' massive particle, with the magnon speed playing the role of the speed of light. In ${\bf m}$-spiral magnets, DW inertia arises from the dimerization mode (i.e. alternating meron displacements along the array). In contrast, meron DWs do not dimerize in ${\bf n}$-spiral magnets. Here, inertia originates from the Berry phase of net magnetization. The dynamics of walls with net topological charge changes drastically in the two types of spiral magnets. While in ${\bf n}$-spiral magnets, it is identical to the zero-charge case, in ${\bf m}$-spiral magnets, a non-local dynamics sets in. That is, the motion of meron DWs is coupled with the precession of spins in the entire spiral domains, hence suppressing wall mobility. An external magnetic field offers fine experimental control over such a non-local dynamics by tuning the net topological charge of the DW. In all the cases described above, hedgehog defects significantly contribute to coercivity through intrinsic pinning.

The results establish spiral multiferroics as a promising platform where magnetoelectric switching and dynamics of topological spin textures are intimately connected. We hope this work motivates experiments on probing meron DWs, confirming different types of defects and manipulating them by external fields. The next step could be to explore how the topological charge (and, thus, magnetization) of single merons can be modified, e.g. by an external magnetic field. The motion of hedgehog defects along meron strings facilitates the topological charge switching. In this perspective, meron domain walls present a promising topologically protected platform in which topological charges provide natural binary variables associated with each meron.

\paragraph{Acknowledgements. ---}
We acknowledge fruitful discussions with M. Fiebig, M. Mostovoy, M. Parodi and F. Foggetti. 
N.N. was supported by JSPS KAKENHI Grant Numbers 24H00197 and 24H02231. N.N. was supported by the RIKEN TRIP initiative.

%

\onecolumngrid
\clearpage

\begin{center}
\textbf{\large Supplementary Information}
\end{center}

\makeatletter

\setcounter{equation}{0}
\setcounter{figure}{0}
\setcounter{table}{0}

\renewcommand{\theequation}{S\arabic{equation}}
\renewcommand{\thefigure}{S\arabic{figure}}
\renewcommand{\thetable}{S\arabic{table}}

\renewcommand{\theHequation}{S\arabic{equation}}
\renewcommand{\theHfigure}{S\arabic{figure}}
\renewcommand{\theHtable}{S\arabic{table}}

\makeatother

\section{Meron domain wall structure: Energy minimization}
Here we report the derivation of \eqref{eq:Theta}. Substituting the configuration \eqref{eq:Ansatz} into the Hamiltonian density \eqref{eq:H0} we obtain
\begin{equation}
\label{eq:HTheta}
\mathcal{H} = \left(K_z \sin^2\!\theta - \frac{1}{2} J_\perp (\partial_y\theta)^2 \right) \sin^2(Qx + \phi_t) - \frac{1}{2} J_x Q^4 ,
\end{equation}
where the constant contribution with $J_x$ is the lowest possible. Indeed, \eqref{eq:Ansatz} minimizes the first term of the Hamiltonian density \eqref{eq:H0} by definition. Requiring $\delta H/\delta\theta = 0$, where $H = \int{dV \: \mathcal{H}}$ and $\delta \cdot /\delta\theta$ denotes the functional derivative with respect to $\theta$, we get a stationary sine-Gordon equation for $\theta(y-\bar{y})$
\begin{equation}
\label{eq:Theta1}
J_\perp \partial_{y}^2\theta + K_z \sin(2\theta) = 0 ,
\end{equation}
where the boundary conditions are: $\theta \rightarrow 0$ for $y-\bar{y} \rightarrow -\infty$ and $\theta \rightarrow \pm\pi$ for $y-\bar{y} \rightarrow +\infty$ (i.e. $\chi = +1$ for $y \ll \bar{y}$ and $\chi = -1$ for $y \gg \bar{y}$). Multiplying by $\partial_y\theta$, \eqref{eq:Theta1} can be rewritten as
\begin{equation}
\label{eq:Theta2}
\frac{1}{2} J_\perp (\partial_y\theta)^2 + K_z \sin^2\!\theta = c ,
\end{equation}
where the boundary conditions imply $c = 0$. Hence, the equation for $\theta(y-\bar{y})$ takes the form
\begin{equation}
\label{eq:Theta3}
\lambda_0^2 (\partial_y\theta)^2 = \sin^2\!\theta, \qquad\qquad \lambda_0 = \sqrt{\frac{\lvert J_\perp \rvert}{2 K_z}} .
\end{equation}
The solutions satisfying the boundary conditions are
\begin{equation}
\label{eq:Theta4}
\theta_\pm(y - \bar{y}) = \pm \arccos\!\left(\!\tanh\!\left(\! -\frac{y-\bar{y}}{\lambda_0} \right)\!\right) ,
\end{equation}
where we required $\theta_\pm(0) = \pm \pi/2$ (i.e. the DW has position $\bar{y}$).

\section{Electric field-driven dynamics: Zero modes}
In this section, we derive $\Gamma_{IJ}$, $G_{IJ}$ and $F_I$ for a meron DW, focusing on the zero modes. Before starting, we briefly review the derivation of the equations of motion for the collective coordinates \eqref{eq:Thiele}. Substituting ${\bf m}(t,{\bf r}) = {\bf m}(\{\xi_I(t)\},{\bf r})$ into \eqref{eq:LeR} and integrating in space and time, one obtains the effective action functional $S$ and dissipation functional $D$
\begin{equation}
\label{eq:LeReff}
S\left[\{\xi_I(t)\}\right] = \int{dt\int{dV \left(\dot{\xi}_J {\bf A}({\bf m}) \cdot \frac{\partial {\bf m}}{\partial \xi_J} - \mathcal{H}_\mathrm{tot}\right)}} , \qquad\qquad D\left[\{\xi_I(t)\}\right] = \frac{\alpha}{2} \int{dt\int{dV \left(\dot{\xi}_J \frac{\partial {\bf m}}{\partial \xi_J}\right)^{\!2}}} , 
\end{equation}
where the dynamical variables are the collective coordinates $\xi_I(t)$. Requiring $\delta S / \delta \xi_I = \delta D / \delta \dot{\xi}_I$, one gets the equations of motion \eqref{eq:Thiele}: $\alpha \Gamma_{IJ} \dot{\xi}_J - G_{IJ} \dot{\xi}_J = F_I$, where \cite{Tretiakov08, Clarke08}
\begin{eqnarray}
\label{eq:Coeff}
\Gamma_{IJ} = \int{dV \: \frac{\partial {\bf m}}{\partial \xi_I} \cdot \frac{\partial {\bf m}}{\partial \xi_J}} , \qquad\quad G_{IJ} = \int{dV \: {\bf m} \cdot \left[ \frac{\partial {\bf m}}{\partial \xi_I} \times \frac{\partial {\bf m}}{\partial \xi_J}\right]} , \qquad\quad F_I = - \int{dV \: \frac{\partial \mathcal{H}_\mathrm{tot}}{\partial \xi_I}} = - \frac{\partial H_\mathrm{tot}}{\partial \xi_I} . 
\end{eqnarray}
If $\xi_I(t)$ is a zero mode, $\partial \mathcal{H} / \partial \xi_I = 0$ by definition and $F_I$ receive contributions only from $\mathcal{H}_\mathrm{DM}$ \eqref{eq:H'}. In the presence of an electric field $E_y$, the DM energy density of the two spiral domains reads $\mathcal{H}_\mathrm{DM} = -\chi \gamma E_y Q$. Hence, a shift of the DW position $\Delta\bar{y}$ corresponds to $\Delta H_\mathrm{DM} = -2 \gamma E_y Q L_x L_z \Delta\bar{y}$, and the conservative force acting on $\bar{y}$ is $F_{\bar{y}} = 2 \gamma E_y Q L_x L_z$. Instead, a rotation of ${\bf m}$ in the $xy$-plane or a translation in the $x$-direction does not change the size of the two spiral domains. Therefore, $\partial H_\mathrm{DM} / \partial \phi_s = \partial H_\mathrm{DM} / \partial \phi_t = 0$ and $E_y$ exerts no force on $\phi_s$ and $\phi_t$.

We observe that $\partial_{\phi_t} = (1/Q)\partial_x$ and $\partial_{\bar{y}} = -\partial_y$. Consequently, $G_{\bar{y} \phi_t}$ can be rewritten as $G_{\bar{y} \phi_t} = 4 \pi L_z \mathcal{Q}_\mathrm{tot} / Q$, where
\begin{equation}
\label{eq:TopQtot}
\mathcal{Q}_\mathrm{tot} = \frac{1}{4\pi} \int{dx dy \: {\bf m} \cdot \left[\partial_x {\bf m} \times \partial_y {\bf m}\right]} = \sum_i \mathcal{Q}_i
\end{equation}
is the total topological charge of the meron DW and $\mathcal{Q}_i$ denotes the topological charge of the $i$-th meron.

Since $\Gamma_{IJ}$ and $G_{IJ}$ are invariant under rotations of ${\bf m}$ (see \eqref{eq:Coeff}), they are independent of $\phi_s$ and can be computed for $\phi_s = 0$. In this case, the configuration \eqref{eq:Ansatz} with the positive solution \eqref{eq:Theta4} can be rewritten as
\begin{equation}
\label{eq:AnsatzS}
{\bf m}(x',y') = (-\sin(Qx'), - \sin\!\left(\Delta\theta(y')\right)\cos(Qx'), \cos\!\left(\Delta\theta(y')\right)\cos(Qx')) ,
\end{equation}
where we defined $x' = x-\bar{x} = x + (\phi_t - \pi/2)/Q$, $y' = y - \bar{y}$ and the odd function $\Delta\theta(y') = \theta_+(y') - \pi/2$. We note that $\mathcal{V} = (-\pi/2Q,\pi/2Q) \times (-l_y,l_y) \times (0,L_z)$, where $l_y \sim \lambda_0$, corresponds to a single meron with center $(\bar{x},\bar{y})$. Both the magnetic texture \eqref{eq:AnsatzS} and the DW region $\mathcal{V}$ are invariant under $\{x' \rightarrow -x',\, m_x \rightarrow -m_x\}$ and $\{y' \rightarrow -y',\, m_y \rightarrow -m_y\}$, but such symmetries act on the derivatives in the zero modes as follows:
\begin{equation}
\begin{aligned}
\label{eq:Sym}
\{x' \rightarrow -x',\, m_x \rightarrow -m_x\} &: \qquad \partial_{\phi_s} \rightarrow -\partial_{\phi_s}, &&\qquad \partial_{\phi_t} \rightarrow -\partial_{\phi_t}, &&\qquad \partial_{\bar{y}} \rightarrow \partial_{\bar{y}}. \\
\{y' \rightarrow -y',\, m_y \rightarrow -m_y\} &: \qquad \partial_{\phi_s} \rightarrow -\partial_{\phi_s}, &&\qquad \partial_{\phi_t} \rightarrow \partial_{\phi_t}, &&\qquad \partial_{\bar{y}} \rightarrow -\partial_{\bar{y}}.
\end{aligned}
\end{equation}
Therefore, every single meron gives no contributions to $\Gamma_{\bar{y}\phi_s}$, $\Gamma_{\bar{y}\phi_t}$ $G_{\bar{y}\phi_s}$, $\Gamma_{\phi_t\phi_s}$ and $G_{\phi_t\phi_s}$. Moreover, $\partial_{\bar{y}} {\bf m}$ is localized within the DW and the spiral domains also give no contributions to the first three. Thus, $\Gamma_{\bar{y}\phi_s}$, $\Gamma_{\bar{y}\phi_t}$ and $G_{\bar{y}\phi_s}$ vanish.

Assuming $\lambda_0 \ll L_y$, the DW gives negligible contributions to $\Gamma_{\phi_s\phi_s}$, $\Gamma_{\phi_t\phi_t}$, $\Gamma_{\phi_t\phi_s}$ and $G_{\phi_t\phi_s}$. Therefore, we compute these components considering only the spiral domains. For $y - \bar{y} \ll -\lambda_0$ and $y - \bar{y} \gg \lambda_0$, the configuration \eqref{eq:Ansatz} reduces to the positive and negative chirality spirals \eqref{eq:Cycloids} with $\phi \equiv \phi_+ = \phi_t +\phi_s$ and $\phi \equiv \phi_- = \phi_t - \phi_s$. Thus, we get
\begin{equation}
\label{eq:CoeffPhi}
y \in (0,\bar{y}): \;\; \partial_{\phi_s}{\bf m} = \partial_{\phi_t}{\bf m}, \qquad\qquad y \in (\bar{y},L_y): \;\; \partial_{\phi_s}{\bf m} = -\partial_{\phi_t}{\bf m}, \qquad\qquad (\partial_{\phi_s}{\bf m})^2 = (\partial_{\phi_t}{\bf m})^2 = 1.
\end{equation}
Combining \eqref{eq:Coeff} and \eqref{eq:CoeffPhi}, we obtain $\Gamma_{\phi_s\phi_s} = \Gamma_{\phi_t\phi_t} = L_x L_y L_z$, $\Gamma_{\phi_t\phi_s} = (2\bar{y} - L_y) L_x L_z$ and $G_{\phi_t\phi_s} = 0$.

To summarize the previous results, $\Gamma_{IJ}$, $G_{IJ}$ and $F_I$ take the form ($I,J = \phi_s, \phi_t, \bar{y}$)
\begin{equation}
\label{eq:Coeff0}
\begin{aligned}
\Gamma_{IJ} = \begin{pmatrix} L_x L_y & L_x (2\bar{y} - L_y) & 0 \\ L_x (2\bar{y} - L_y) & L_x L_y & 0 \\ 0 & 0 & \Gamma_{\bar{y}\bar{y}} \end{pmatrix} \! L_z , \quad
G_{IJ} = \begin{pmatrix} \, 0 & 0 & 0 \\ \,  0 & 0 & -4 \pi \mathcal{Q}_\mathrm{tot} / Q \\ \, 0 & 4 \pi \mathcal{Q}_\mathrm{tot} / Q & 0 \end{pmatrix} \! L_z , \quad
F_I = \begin{pmatrix} 0 \\ 0 \\ 2 \gamma E_y Q L_x \end{pmatrix} \! L_z ,
\end{aligned}
\end{equation}
where $\Gamma_{\bar{y}\bar{y}}$ is the only component that depends on the internal structure of the meron DW. Using \eqref{eq:Ansatz} and \eqref{eq:Theta}, we get $\Gamma_{\bar{y}\bar{y}} = L_x L_z / \lambda_0$. If the DW width $\lambda$ is a function of $x$, we obtain instead
\begin{equation}
\label{eq:GammaDef}
\Gamma'_{\bar{y}\bar{y}} = L_z \int_0^{L_x}\!{dx \: \frac{2 \sin^2(Qx + \phi_t)}{\lambda(x)}} \approx L_z \int_0^{L_x}\!{\frac{dx}{\lambda(x)}},
\end{equation}
where we assumed that $\lambda(x)$ is nearly constant in an interval $\Delta x = \pi/Q$ to replace $2 \sin^2(Qx + \phi_t)$ with the average.

Using these results, we obtain the long-time dynamics \eqref{eq:Steady} and \eqref{eq:SolDef}. At the first order in $E_y$, the hard modes do not affect such a dynamics. Let us consider a generic hard mode $h$ in the case of \eqref{eq:Steady}, that is, meron DWs with alternating $\mathcal{Q}_i$. In the steady motion regime, $\dot{h} = 0$ and therefore the equation of motion \eqref{eq:Thiele} for $\bar{y}$ is still $\alpha \Gamma_{\bar{y}\bar{y}} \dot{\bar{y}} = F_{\bar{y}}$. However, $h$ stays in its equilibrium position $h_{\infty}(E_y)$ and this corresponds to a deformation of the configuration \eqref{eq:Ansatz}. Thus, $\Gamma_{\bar{y}\bar{y}}$ and $ F_{\bar{y}}$ may be modified. Nevertheless, the contribution to the previous equation is at least a second-order correction since $h_{\infty}(E_y)$ is at least linear in $E_y$ (i.e. $h_{\infty}(0) = 0$) and the equation has no zero-order term. Indeed, $\dot{\bar{y}}$ and $F_{\bar{y}}$ are both linear in $E_y$.

\section{Electric field-driven dynamics: Dimerization and breathing modes}
As discussed in the main text, the dimerized meron DW at time $t$ takes the form
\begin{equation}
\label{eq:FullAnsatz}
{\bf m}(t,x,y) = \hat{\mathcal{R}}_{\hat{z}}(\phi_s) \, \hat{\mathcal{R}}_{\hat{x}} \! \left(\theta(y - \bar{y}(t))\right) \begin{pmatrix} \cos(Qx + \phi_t + f(t, x, y - \bar{y}(t))) \\ \sin(Qx + \phi_t + f(t, x, y - \bar{y}(t))) \\ 0 \end{pmatrix}.
\end{equation}
In the steady motion regime, $f$ has no explicit time dependence and $\bar{y}(t) = v_{\infty}t$ (i.e. $f(t, x, y - \bar{y}(t)) \equiv f_\infty(x, y - v_{\infty}t)$). Substituting \eqref{eq:FullAnsatz} into \eqref{eq:LeR} and requiring $\delta S / \delta f_\infty = \delta D / \delta \dot{f}_\infty$, where $S\left[f_\infty\right] = \int{dt \int{dV \: \mathcal{L}}}$ and $D\left[f_\infty\right] = \int{dt \int{dV \: \mathcal{R}}}$, we get the following equation for $f_\infty$ at the first order in $v_{\infty}$
\begin{equation}
\label{eq:Eqf}
\partial_x^4 f_\infty(x, y - v_{\infty}t) - 4 Q^2 \partial_x^2 f_\infty(x, y - v_{\infty}t) = \frac{v_{\infty}}{J_x \lambda_0} \frac{\sin(Qx + \phi_t)}{\cosh({(y - v_{\infty}t)/\lambda_0})} ,
\end{equation}
where we used $K_z  \ll J_x$.
Seeking a solution of the form $f_\infty(x, y - v_{\infty}t) = \sin(Qx + \phi_t) \, g(y - v_\infty t)$, we get \eqref{eq:f}. Then, promoting $d_\infty$ to $d(t)$, $f$ takes the form
\begin{equation}
\label{eq:fGen}
f(t, x, y - \bar{y}(t)) = d(t) \frac{\sin(Qx + \phi_t)}{\cosh({(y - \bar{y}(t))/\lambda_0})} .
\end{equation}
Above, we defined the dimerized configuration in the transient from that in the steady motion regime. It is possible to give an alternative (but equivalent) definition, based on the action of $E_y$ at the initial time $t=0$. The magnetization field ${\bf m}(t,{\bf r})$ obeys the LLG equation \cite{Landau35, Gilbert04}, obtained from \eqref{eq:LeR} as ${\bf m} \times (\delta S / \delta {\bf m} - \delta D / \delta \dot{{\bf m}}) = 0$,
\begin{equation}
\label{eq:LLG}
\dot{{\bf m}} = {\bf m} \times \frac{\delta H_\mathrm{tot}}{\delta {\bf m}} + \alpha \, {\bf m} \times \dot{{\bf m}} .
\end{equation}
In particular, the contribution of the electric field $E_y$ reads $(\dot{{\bf m}})_\mathrm{DM} = {\bf m} \times \delta H_\mathrm{DM} / \delta {\bf m} = 2 \gamma E_y \, {\bf m} \times[\hat{z} \times \partial_x {\bf m}]$. After a short time interval $\Delta t$, the meron DW is deformed: ${\bf m}(\Delta t,{\bf r}) \approx {\bf m}(0,{\bf r}) + (\dot{{\bf m}})_\mathrm{DM} \Delta t$, where we neglected the dissipation. Using the static magnetization texture \eqref{eq:Ansatz} (i.e. the configuration at the initial time), we get $(\dot{{\bf m}})_\mathrm{DM} \cdot \partial_{\bar{y}}{\bf m} = 0$. Hence, such a deformation does not involve $\bar{y}$ but another collective coordinate $d$ so that
\begin{equation}
\label{eq:FullAnsatzLLG}
{\bf m}(t,x,y) = {\bf m}(0,x,y) - \frac{d(t)}{Q} \, {\bf m}(0,x,y) \times\left[\hat{z} \times \partial_x {\bf m}(0,x,y)\right],
\end{equation}
where the factor $-1/Q$ is added for convenience, and ${\bf m}(0,x,y)$ is given by \eqref{eq:Ansatz} and \eqref{eq:Theta}. The previous equation defines the dimerized configuration at the first order in $d$ and, consistently, it coincides with the first order expansion of \eqref{eq:FullAnsatz} and \eqref{eq:fGen}. The fact that the driving force does not act directly on the DW position but only changes $d$ corroborates the identification of the dimerization mode with the momentum conjugate to $\bar{y}$.

To derive \eqref{eq:FullAnsatz} and \eqref{eq:fGen}, we did not use the explicit forms of $\lambda_0$ and $v_\infty$. Thus, these are true even at high electric fields, when $\lambda_0 \rightarrow \lambda(t)$ and the steady velocity differs from \eqref{eq:Steady}. Up to the second order in $d$, \eqref{eq:Thiele} takes the form 
\begin{equation}
\label{eq:EqYDL1}
\frac{\alpha}{\lambda_0} \begin{pmatrix} \gamma_c & 0 & 0 \\ 0 & \lambda_0^2/\gamma_c & d \lambda_0/2 \\ 0 & d \lambda_0/2 & \gamma_c \pi^2 / 12 \end{pmatrix} \begin{pmatrix} \dot{\bar{y}} \\ \dot{d} \\ \dot{\lambda} \end{pmatrix} - \begin{pmatrix} 0 & -1 & 0 \\ 1 & 0 & 0 \\ 0 & 0 & 0 \end{pmatrix}  \begin{pmatrix} \dot{\bar{y}} \\ \dot{d} \\ \dot{\lambda} \end{pmatrix} = \begin{pmatrix} q E_y \\ - d/m \gamma_c \\ - K_z \left(1 - \gamma_c^2 + d^2 / m^2 c^2_\mathrm{m} \right) \end{pmatrix},
\end{equation}
where $\gamma_c = \lambda_0/\lambda$, $q = 2 Q \gamma$, $m^{-1} = 5 J_x \lambda_0 Q^4$, $c_\mathrm{m} = \sqrt{5 \lvert J_\perp \rvert J_x Q^4}$, and the second-order corrections of $f(t, x, y - \bar{y}(t))$ do not contribute for $K_z \ll J_x$.
Neglecting second-order terms in $\alpha$, $\dot{d}$ and $\dot{\lambda}$, \eqref{eq:EqYDL1} becomes
\begin{equation}
\label{eq:EqYDL2}
\dot{d} = q E_y - \beta \gamma_c \dot{\bar{y}}, \qquad d = m \gamma_c \dot{\bar{y}}, \qquad \gamma_c = \frac{1}{\sqrt{1 - \dot{\bar{y}}^2 / c^2_\mathrm{m}}},
\end{equation}
where $\beta = \alpha/\lambda_0$, and $\gamma_c$ resembles the Lorentz factor with $c_\mathrm{m}$ playing the role of the speed of light.

\section{Defects within the meron wall: Dynamics of multiple domain walls}
Let us consider $K$ meron DWs with positions $\bar{y}_k$, which separate $K+1$ spiral domains with phases $\phi_k$ and alternating chiralities $\chi_k$. For $I = \phi_k$, the equations of motion \eqref{eq:Thiele} express the precession speed of ${\bf m}$ in the $k$-th domain in terms of the velocities of the adjacent DWs:
\begin{equation}
\label{eq:Phases}
\dot{\phi}_k = -\frac{2}{\alpha N} \frac{\mathcal{Q}_k \dot{\bar{y}}_k + \mathcal{Q}_{k-1} \dot{\bar{y}}_{k-1}}{\bar{y}_k - \bar{y}_{k-1}},
\end{equation}
where $N = L_x Q / \pi$ is the total number of merons within a single DW, $\mathcal{Q}_k$ denotes the total topological charge of the $k$-th wall, $\bar{y}_0 \equiv 0$, $\bar{y}_{K+1} \equiv L_y$ and $\dot{\bar{y}}_0 = \dot{\bar{y}}_{K+1} \equiv 0$ are defined to simplify the notation. For $I = \bar{y}_k$ with $k = 1,2,\ldots,K$, the equations of motion \eqref{eq:Thiele} thus take the form
\begingroup
\renewcommand*{\arraystretch}{1.35}
\begin{equation}
\label{eq:Positions}
\begin{aligned}
\begin{pmatrix} C_{11} & C_{12} & 0 & 0 & \cdots \\ C_{12} & C_{22} & C_{23} & 0 & \cdots \\ 0 & C_{23} & C_{33} & C_{34} & \cdots \\ 0 & 0 & C_{34} & C_{44} & \ddots \\ \vdots & \vdots & \vdots & \ddots & \ddots \end{pmatrix} \begin{pmatrix} \dot{\bar{y}}_1 \\ \dot{\bar{y}}_2 \\ \dot{\bar{y}}_3 \\ \dot{\bar{y}}_4 \\ \vdots \end{pmatrix} = \begin{pmatrix} \pm v_\infty \\ \mp v_\infty \\ \pm v_\infty \\ \mp v_\infty \\ \vdots \end{pmatrix},
\end{aligned}
\end{equation}
\endgroup
where upper and lower signs in front of $v_\infty$ \eqref{eq:Steady} correspond to the cases in which the first domain has chirality $\chi_1 = +1$ and $\chi_1 = -1$, respectively, and
\begin{equation}
\label{eq:C}
C_{kk} = b_k + \frac{\lambda_0}{\alpha^2} \frac{\bar{y}_{k+1} - \bar{y}_{k-1}}{(\bar{y}_k - \bar{y}_{k-1})(\bar{y}_{k+1} - \bar{y}_k)} \left(\frac{2 \mathcal{Q}_k}{N}\right)^2 , \qquad C_{kl} = C_{lk} = \frac{\lambda_0}{\alpha^2} \frac{1}{\lvert \bar{y}_k - \bar{y}_l \rvert} \frac{4 \mathcal{Q}_k \mathcal{Q}_l}{N^2} \quad \mathrm{for} \ \lvert k-l \rvert = 1 .
\end{equation}
The $k$-th DW exerts a force $F_{kl} \propto C_{kl} \dot{\bar{y}}_k$ on the neighboring walls (labeled by $l$). Since $C_{kl}$ is proportional to $\mathcal{Q}_k$ and $\mathcal{Q}_l$, a meron DW with zero total topological charge does not interact with the others.

\section{Atomistic spin dynamics simulations: Minimum energy meron domain wall}
To support the analytical results, we perform various atomistic spin dynamics simulations using the UppASD code \cite{Skubic08}, which numerically solves the LLG equation \eqref{eq:LLG}. We first consider a simplified lattice model,
\begin{equation}
\label{eq:Disc}
H = \sum_{\bf r} \left(J_1 \, {\bf S}_{\bf r} \cdot {\bf S}_{{\bf r} + \hat{x}} + J_2 \, {\bf S}_{\bf r} \cdot {\bf S}_{{\bf r} + 2\hat{x}} + J_\perp ({\bf S}_{\bf r} \cdot {\bf S}_{{\bf r} + \hat{y}} + {\bf S}_{\bf r} \cdot {\bf S}_{{\bf r} + \hat{z}}) + K_z \left({\bf S}_{\bf r} \cdot \hat{z}\right)^2 - \gamma E_y \hat{z} \cdot \left[{\bf S}_{\bf r} \times {\bf S}_{{\bf r} + \hat{x}}\right]\right),
\end{equation}
with competing nearest neighbor $J_1<0$ and next to nearest neighbor $J_2>0$ exchange interactions in the $x$-direction, ferromagnetic $J_\perp<0$ in the $y$ and $z$ directions, easy $xy$-plane anisotropy $K_z>0$ and DM vector $- \gamma E_y \hat{z}$. For ${\pi/Q\gg 1}$ and ${\lambda_0 \gg 1}$, the Hamiltonian \eqref{eq:Disc} is equivalent to the spatial integral of $\mathcal{H}_\mathrm{tot}$ (see \eqref{eq:H0} and \eqref{eq:H'}) with $J_x = J_2 \sin^2\!Q/Q^2$ and $Q = \arccos(-J_1/4 J_2)$.

\begin{figure}[b]
\centering
\includegraphics[width=1\linewidth]{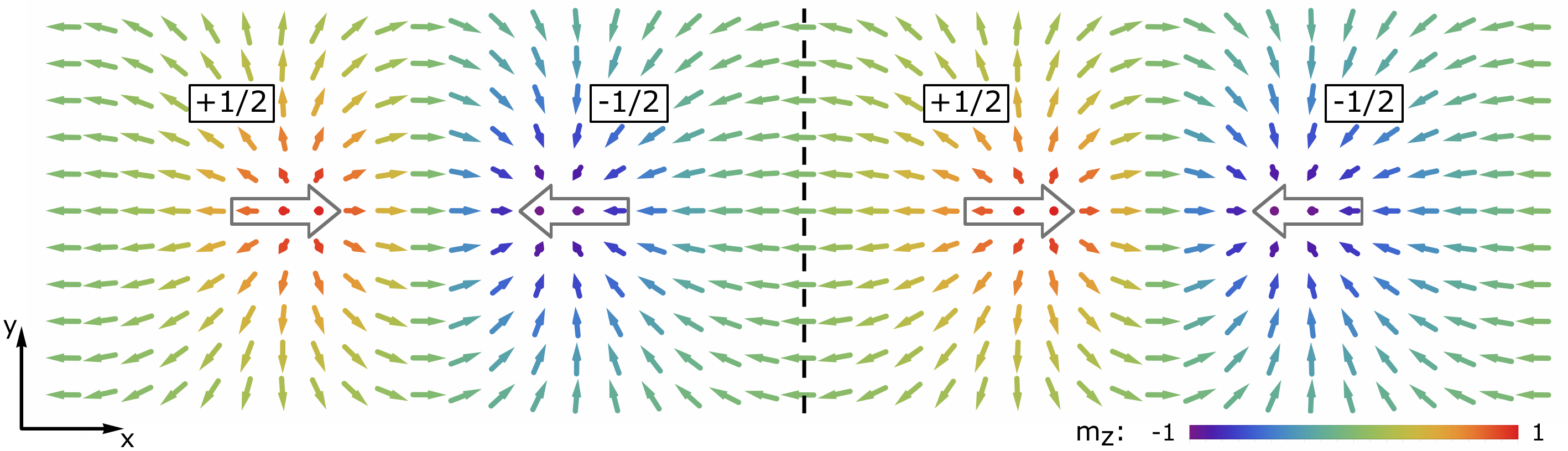}
\caption{\label{fig:DimFit}Meron DW moving under the action of an applied electric field $E_y$. Gray arrows highlight dimerization (i.e. alternating meron displacements). The dashed line separates the magnetization texture observed in the simulation (left) and the analytical solution with fitted $\phi_s$, $\phi_t$, $\bar{y}$ and $d$ (right).
}\end{figure}

\begin{figure}[p]
\centering
\includegraphics[width=1\linewidth]{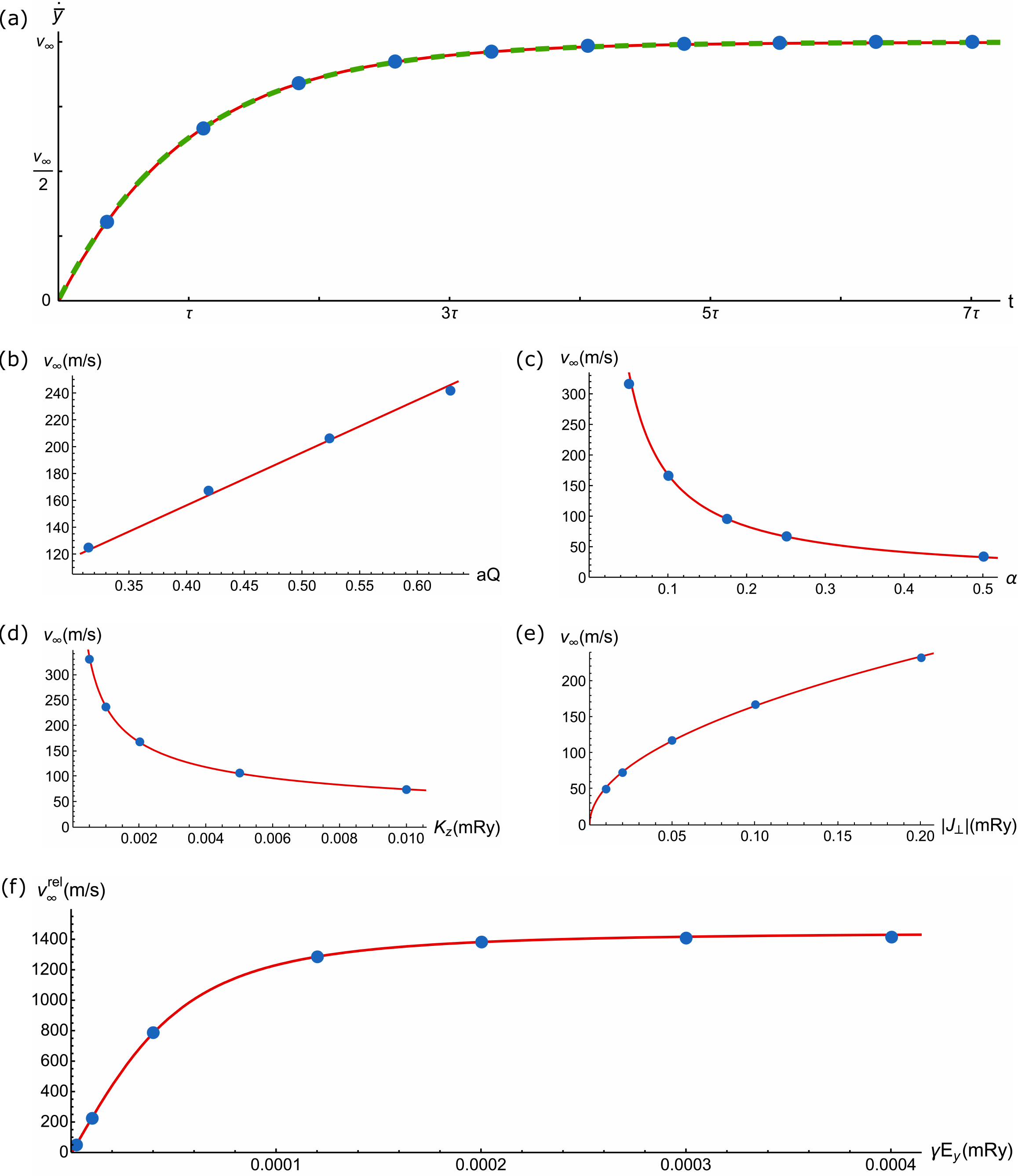}
\caption{\label{fig:VelMin}(a) Meron DW velocity as a function of time. The green and red lines represent the numerical result and the analytical solution \eqref{eq:Transient} with fitted $v_\infty$ and $\tau$ (here $v_0 = 0$). The Blue dots indicate $d/m$ as a function of $t$. The dimerization $d$ is obtained by fitting the configuration given by \eqref{eq:FullAnsatz}, \eqref{eq:Theta4} and \eqref{eq:fGen} to the magnetization texture observed in simulations. The effective mass $m$ is determined by averaging the numerical $d/\dot{\bar{y}}$ (see \eqref{eq:EqYD2}). Panels (b), (c), (d), (e) and (f) show the DW steady velocity as a function of $a Q$, $\alpha$, $K_z$, $\lvert J_\perp \rvert$ and $\gamma E_y$, respectively. The blue dots indicate the numerical results, and the red lines represent the fitted analytical solutions \eqref{eq:Steady} and \eqref{eq:SteadyRel}.
}\end{figure}

We simulate a $60 \times 90 \times 1$ system with periodic boundary conditions in the $x$-direction and open boundary conditions in the $y$-direction. The magnetic moment magnitude of a site is set to the Bohr magneton $\mu_\mathrm{B}$. In the first simulation, we choose $J_1 = -1.4 \; \mathrm{mRy}$, $J_2 = 0.38 \; \mathrm{mRy}$, $J_\perp = -0.1 \; \mathrm{mRy}$, $K_z = 0.002 \; \mathrm{mRy}$, $\gamma E_y = 0.0002 \; \mathrm{mRy}$, $\alpha = 0.1$, the time step $\delta t = 3 \cdot 10^{-16} \, \mathrm{s}$ and the total number of iterations $\mathcal{N} = 6 \cdot 10^{5}$. The initial state is the minimum energy DW \eqref{eq:Ansatz} corresponding to $J_1$, $J_2$, $J_\perp$ and $K_z$. Restoring the spin $S = \hbar/2$ and the lattice constant $a = 0.5\;\mathrm{nm}$ we get
\begin{equation}
\begin{aligned}
\label{eq:Sim}
\mathrm{Analytical:}& \qquad v_\infty = 173 \; \mathrm{m/s}, && \qquad \tau = 4.35 \cdot 10^{-12} \, \mathrm{s}, && \qquad d_\infty = 0.030, \\
\mathrm{Simulation:}& \qquad v_\infty = 167 \; \mathrm{m/s}, && \qquad \tau = 4.89 \cdot 10^{-12} \, \mathrm{s}, && \qquad d_\infty = 0.033,
\end{aligned}
\end{equation}
where $\tau$ denotes the relaxation time of the dimerization mode, $v_\infty$ and $d_\infty$ are the steady-state velocity and dimerization. The comparison indicates a good agreement between the analytical results and the simulations, with the relative errors of $3.6 \%$, $12 \%$ and $10 \%$, respectively. To further corroborate the analytical model, we change $J_2$, $J_\perp$, $K_z$ and $\alpha$ one by one. Figure~\ref{fig:VelMin} shows the results of these additional simulations, as well as the fit of $\dot{\bar{y}}(t)$ \eqref{eq:Transient} that determines $v_\infty$ and $\tau$. Instead, $d_\infty$ is obtained by fitting the magnetization texture observed in simulations by the ansatz \eqref{eq:FullAnsatz}, \eqref{eq:Theta4} and \eqref{eq:fGen} (see Fig.~\ref{fig:DimFit}). In order to better visualize dimerization in Fig.~\ref{fig:DimFit} and Movie~1, we enhance it by using a different set of parameters: $J_1 = -1.52 \; \mathrm{mRy}$, $J_2 = 0.4 \; \mathrm{mRy}$, $J_\perp = -0.1 \; \mathrm{mRy}$, $K_z = 0.004 \; \mathrm{mRy}$, $\gamma E_y = 0.002 \; \mathrm{mRy}$. Still, the dimerization from analytical and numerical calculations, $d_\infty = 0.67$ and $d_\infty = 0.62$, are in good agreement.

We repeat the simulation \eqref{eq:Sim} using open (instead of periodic) boundary conditions in the $x$-direction, which may be a more realistic setup. The initial configuration is numerically relaxed before the electric field $E_y$ is applied. The fit of $\dot{\bar{y}}(t)$ \eqref{eq:Transient} is still good, and the resulting steady velocity and transient time are $v_\infty = 149 \; \mathrm{m/s}$ and $\tau = 5.57 \cdot 10^{-12} \, \mathrm{s}$. Since the magnetization texture is deformed near the surfaces, we restrict the fit of \eqref{eq:FullAnsatz} to the central region of the DW, obtaining $d_\infty = 0.034$. Increasing $L_x$ from $60$ to $150$, we get  $v_\infty = 161 \; \mathrm{m/s}$, $\tau = 5.21 \cdot 10^{-12} \, \mathrm{s}$ and $d_\infty = 0.034$, which are in better agreement with \eqref{eq:Sim}. The larger the system, the less relevant the surface effects are. Hence, for $L_x \gg \pi/Q$, we expect to recover \eqref{eq:Sim}.

To test the ``relativistic'' DW dynamics \eqref{eq:EqRel} and \eqref{eq:SteadyRel}, we run several simulations of a $60 \times 500 \times 1$ system with periodic boundary conditions in the $x$-direction, open boundary conditions in the $y$-direction and parameters: $J_1 = -1.4 \; \mathrm{mRy}$, $J_2 = 0.38 \; \mathrm{mRy}$, $J_\perp = -0.1 \; \mathrm{mRy}$, $K_z = 0.001 \; \mathrm{mRy}$, $\alpha = 0.005$ and $\gamma E_y \in (2.5 \cdot 10^{-6}, \, 4 \cdot 10^{-4}) \, \mathrm{mRy}$. We obtain
\begin{equation}
\begin{aligned}
\label{eq:SimRel}
\mathrm{Analytical:}& \qquad c_\mathrm{m} = 1541 \; \mathrm{m/s}, && \qquad \mu_\mathrm{DW} = 2.45 \cdot 10^7 \, \mathrm{m/(s\,mRy)}, \\
\mathrm{Simulations:}& \qquad c_\mathrm{m} = 1450 \; \mathrm{m/s}, && \qquad \mu_\mathrm{DW} = 2.36 \cdot 10^7 \, \mathrm{m/(s\,mRy)},
\end{aligned}
\end{equation}
where $c_\mathrm{m}$ is the limiting velocity, and $\mu_\mathrm{DW} = v_\infty / \gamma E_y$ denotes the low-field DW mobility. The relative errors are $6.1 \%$ and $3.7 \%$, respectively. Figure~\ref{fig:VelMin}(f) shows the fit of \eqref{eq:SteadyRel} that determines \eqref{eq:SimRel}. With open boundary conditions in the $x$-direction, defects (see Fig.~\ref{fig:Defect}) are nucleated at the ends of the wall when the DW velocity approaches $c_\mathrm{m}$.

\section{Atomistic spin dynamics simulations: Defects within the meron wall}
\begin{figure}[p]
\centering
\includegraphics[width=1\linewidth]{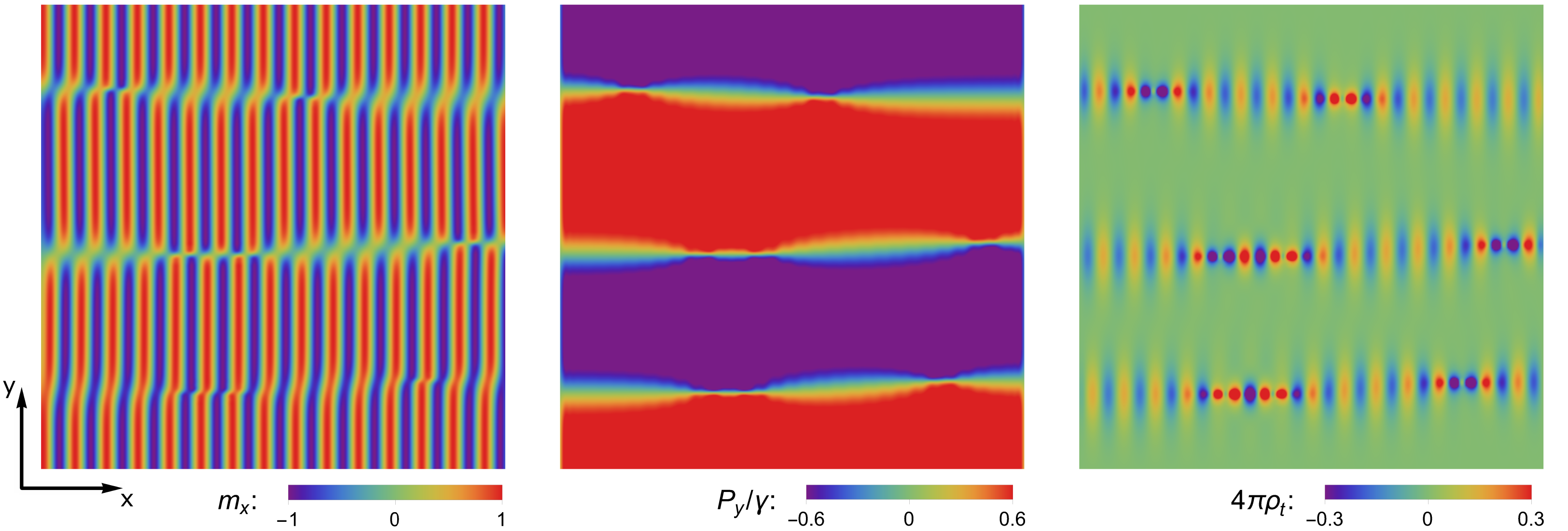}
\caption{\label{fig:Random}Magnetization texture obtained by relaxing a random configuration. From left to right, the density plots represent the magnetization $x$-component $m_x$, the (re-scaled) polarization $y$-component $P_y/\gamma$ and the (re-scaled) topological charge density $4 \pi \rho_t$. Three meron DWs with defects separate four spiral domains. From $P_y/\gamma$ and $4 \pi \rho_t$ it is clear that the walls are narrower near the defects. Furthermore, the two lower DWs have a non-zero total topological charge.
}\end{figure}

To study spontaneous defect formation across the ordering transition, we start the simulation from a random spin configuration (i.e. each spin points in a random direction). We consider a $120 \times 120 \times 1$ system with open boundary conditions, and we set: $J_1 = -1.08 \; \mathrm{mRy}$, $J_2 = 0.38 \; \mathrm{mRy}$, $J_\perp = -0.1 \; \mathrm{mRy}$, $K_z = 0.002 \; \mathrm{mRy}$, $\alpha = 0.1$, $\delta t = 2 \cdot 10^{-16} \, \mathrm{s}$, $\mathcal{N} = 4 \cdot 10^{5}$ and $E_y=0$. The resulting spin texture consists of four spiral domains separated by three meron DWs with defects, two of which have $\mathcal{Q}_\mathrm{tot} \neq 0$ (see Fig.~\ref{fig:Random}). This confirms that defects within the meron chain are metastable and can lead to DWs with a net topological charge $\mathcal{Q}_\mathrm{tot}$. We note that the conditions required to obtain only meron DWs by relaxing a random configuration are stricter than DW metastability. For example, if $J_1 = -1.4 \; \mathrm{mRy}$, meron DWs are metastable (in fact, by relaxing the spin configuration \eqref{eq:Ansatz}, only negligible corrections in $K_z$ appear), but the simulation where a random configuration is relaxed does not result in meron DWs. For this to happen, meron DWs must be much lower in energy than any other DW type. In particular, considering the competition with Hubert DWs, such a condition takes the form $L_x \sqrt{K_z \lvert J_\perp \rvert} \ll L_y J_2 Q \sin^2\!Q$, where $a$ is set to $1$. Therefore, we can compensate for the increase in $\lvert J_1 \rvert$ (i.e. the decrease in $Q$) by increasing $L_y/L_x$.

\begin{figure}[p]
\centering
\includegraphics[width=1\linewidth]{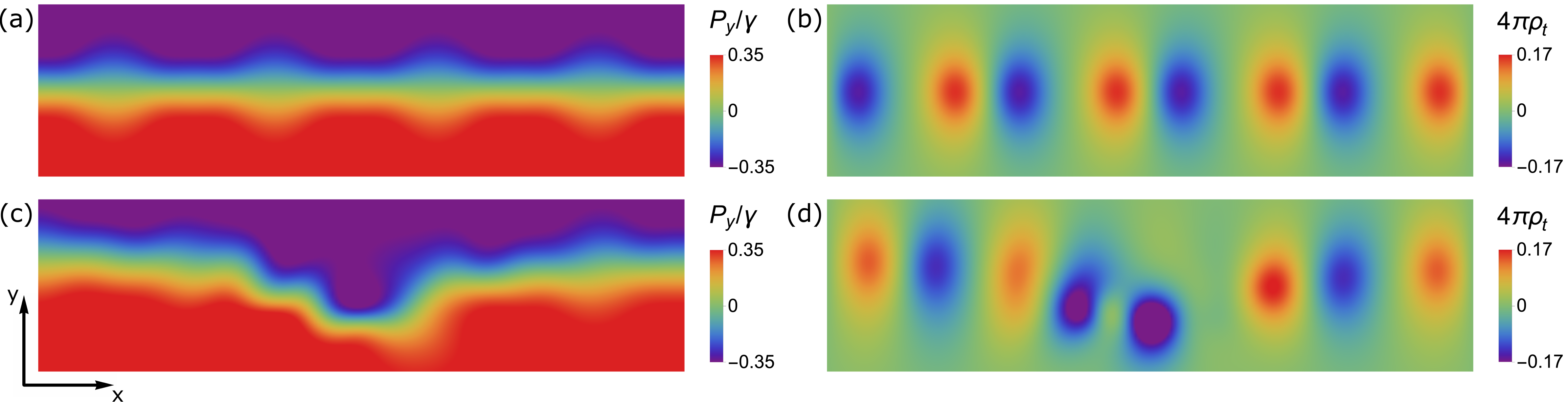}
\caption{\label{fig:DefCyc}Minimum energy DW (a),(b) and defected DW with $\mathcal{Q}_\mathrm{tot} = 0$ (c),(d) moving under the action of $E_y$. The left panels show the (re-scaled) polarization $y$-component $P_y/\gamma$. The right panels represent the (re-scaled) topological charge density $4 \pi \rho_t$. Merons near the defect are falling behind, and the DW is deformed.
}\end{figure}

\begin{figure}[p]
\centering
\includegraphics[width=1\linewidth]{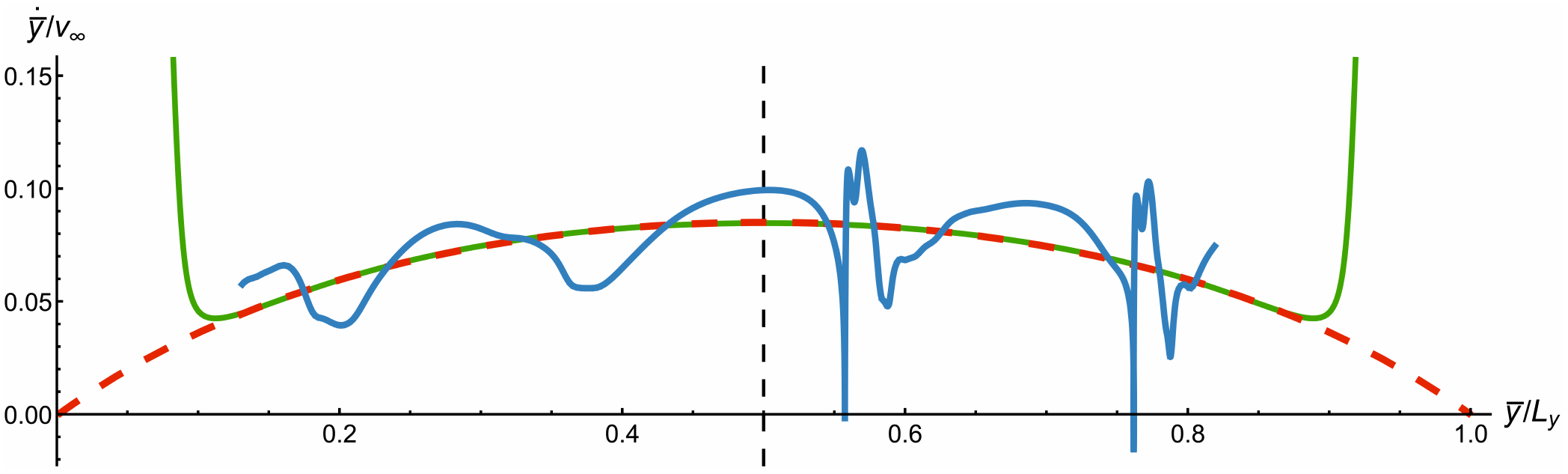}
\caption{\label{fig:VelDef}Velocity of a meron DW with $\mathcal{Q}_\mathrm{tot} \neq 0$ as a function of its position. The green and red lines indicate the numerical result with periodic boundary conditions in the $x$-direction and the analytical solution \eqref{eq:SolDef} with fitted parameters.
Because of open boundary conditions in the $y$-direction, the surface attracts the DW when the distance between the two is comparable to the wall width. Equation~\eqref{eq:SolDef} does not take this effect into account. The blue line represents a simulation with open boundary conditions in the $x$-direction. In the graphic left half, merons with alternating topological charges are destroyed, and $\dot{\bar{y}}$ oscillates around \eqref{eq:SolDef} (red line). In the right half, two defects are destroyed, and $\dot{\bar{y}}$ drops sharply at these events.
}\end{figure}

In order to simulate the dynamics of meron DWs with defects, we proceed similarly to the case of minimum energy DWs. The main difference is that now we can not rely on the analytical expression for the DW structure to create the initial configuration because $\lambda(x)$ is unknown. Hence, we reverse $m_z$ within a few merons in \eqref{eq:Ansatz} and then numerically relax the configuration. As discussed in the main text, the damping force $F_\Gamma^i$ is stronger near the defects. Consistently, in simulations, the merons near the defects are falling behind (see Fig.~\ref{fig:DefCyc}). The collective coordinate corresponding to this deformation is a hard mode. Thus, we neglect it in the derivation of the long-time dynamics \eqref{eq:SolDef}. In addition, we consider a DW where all the merons have the same topological charge (see Fig.~\ref{fig:VelDef} and Movie~2). The simulation
clearly shows bulk rotations of spins and agrees with \eqref{eq:SolDef} far from the surfaces $y = 0$ and $y = L_y$. Still, the analytical model, which does not take surface effects into account, fails to describe the near-surface behavior. Indeed, it predicts that the DW velocity $\dot{\bar{y}}$ goes to zero at the surfaces, while in the simulation $\dot{\bar{y}}$ increases sharply due to open boundary conditions in the $y$-direction. In the above simulation, the boundary conditions in the $x$-direction are periodic. Using open boundary conditions instead, the dynamics changes significantly. Bulk rotations of spins are still observed, but only above a critical electric field. Indeed, the driving field must overcome an energy barrier to push a meron through the boundary and activate $\phi_t$. Furthermore, the destruction of a defect at one end of the wall (see Movie~3) coincides with a drop in the velocity $\dot{\bar{y}}$ (see Fig.~\ref{fig:VelDef}, right side), which is not described by \eqref{eq:SolDef}. Since merons created at the other end have alternating topological charges (see Movie~3), $\mathcal{Q}_\mathrm{tot}$ reduces as the DW moves. Instead, when the destroyed merons have alternating topological charges (see Movie~4), the velocity $\dot{\bar{y}}$ oscillates around \eqref{eq:SolDef} (see Fig.~\ref{fig:VelDef}, left side), and $\mathcal{Q}_\mathrm{tot}$ remains the same. Hence, the analytical solution \eqref{eq:SolDef} captures the averaged DW velocity.

We verify the stability of hedgehog defects by relaxing two $60 \times 60 \times 60$ initial configurations of the types shown in Fig.~\ref{fig:TopControl}(f,$\,$g), using periodic boundary conditions in the $x$-direction. Near hedgehogs, the DW width becomes comparable to the lattice constant. Furthermore, by relaxing a random configuration in a $80 \times 120 \times 60$ system with open boundary conditions, we obtain a meron DW presenting an array of hedgehog defects (see Fig.~\ref{fig:TopControl}(c,$\,$f)). Consequently, we expect these defects to arise naturally during domain nucleation.

\section{\texorpdfstring{Atomistic spin dynamics simulations: Meron domain walls in $\mathbf{TbMnO}_3$}{Atomistic spin dynamics simulations: Meron domain walls in TbMnO3}}
We also simulate meron DW motion in TbMnO$_3$ using the microscopic spin Hamiltonian proposed by Mochizuki et al. (see \cite{Mochizuki09b} for a detailed description of the model). The P$bnm$ orthorhombic unit cell, with axes $a$, $b$ and $c$, contains four manganese ions with spin $S=2$. The nearest neighbor exchange interaction in the $ab\,$-plane $J_{ab}$ is ferromagnetic, while it is antiferromagnetic in the $c\,$-direction, $J_c > 0$. Frustration originates from the antiferromagnetic next-nearest neighbor exchange interaction in the $b\,$-direction, $J_2$. Hence, the spiral wave vector ${\bf Q}$ is parallel to the $b\,$-axis. Since in our model ${\bf Q}$ is oriented along $x$, we choose the axes so that $x = b$, $y = -a$ and $z = c$. Assuming a smooth varying spin texture ${\bf S}({\bf r})$ and expanding the exchange Hamiltonian in powers of spin gradients, we express $J_x$, $Q$ and $J_\perp$ (see \eqref{eq:H0} and \eqref{eq:LeRAf}) in terms of the exchange parameters of TbMnO$_3$:
\begin{equation}
\label{eq:TbMnO3}
Q = \pm 2 \arccos{\!\left(\frac{\lvert J_{ab} \rvert}{2 J_2}\right)}, \qquad J_x = \frac{J_2 S^2}{4 Q^2}\sin^2{\!\left(\frac{Q}{2}\right)}, \qquad J_\perp^{(y)} = \frac{J_{ab} S^2}{2}\cos{\!\left(\frac{Q}{2}\right)}, \qquad J_\perp^{(z)} = J_c S^2,
\end{equation}
where $J_\perp^{(y)}$ and $J_\perp^{(z)}$ enter the terms of the Hamiltonian density that involve derivatives with respect to $y$ and $z$. The kinetic term of the Lagrangian density \eqref{eq:LeRAf} is determined solely by the antiferromagnetic interaction $J_\perp^{(z)}$ (see \cite{Dasgupta21} for a detailed derivation). The model for TbMnO$_3$ also includes single-ion anisotropy along the tilted axes of each MnO$_6$ octahedron and Dzyaloshinskii-Moriya (DM) interactions characterized by opposite DM vectors for neighboring bonds of the same orientation. These terms lead to the effective easy $b\,$-axis anisotropy $K_b < 0$ and easy $bc\,$-plane anisotropy $K_a > 0$. The easy-axis anisotropy does not affect the dynamics of meron DWs in the antiferromagnetic case, since $\phi_s$ is never an active mode. The easy-plane anisotropy enters the results only through the width $\lambda_0$ of a static DW. We measure $\lambda_0$ by relaxing a meron DW at zero applied field. The effective anisotropy can be written as $K_a = \lvert J_\perp \rvert/2\lambda_0^2$.

The ferroelectric polarization ${\bf P}$ induced by the spiral order lies in the spiral plane and is perpendicular to the wave vector. Consequently, ${\bf P}$ is parallel to the $c\,$-axis. We examine the meron DW dynamics under an applied electric field $E$ in this direction, which enters the model through the DM interaction associated with the polar distortions along $c$. Each wall with a normal $\hat{e}_\perp$ lying in the $ac\,$-plane is a meron DW. Nevertheless, experiments indicate that walls with $\hat{e}_\perp = \hat{a}$ are preferred \cite{Matsubara15}, probably because they do not carry any electric charge. Therefore, we simulate the electric field-driven dynamics of a meron DW with this orientation, using the parameters for TbMnO$_3$ introduced in \cite{Mochizuki10c, Matsubara15} and the magnetoelectric coupling constant $\gamma$ estimated from the experimental data \cite{Kimura03}. We follow a similar approach as in the previous case to verify the structure of the meron DW \eqref{eq:Ansatz}, the stability of defects (see Fig.~\ref{fig:TopControl}(e--g)), and the dynamics described by \eqref{eq:EqRel}, with the replacement $5 J_x Q^4 \rightarrow 4 J_\perp^{(z)} N_\mathrm{AF}$ discussed in the main text. $J_\perp$ corresponds to $J_\perp^{(y)}$ when it does not descend from this replacement. Within the range of fields commonly applied in experiments, the DW steady-state velocity is directly proportional to the electric field: $v_\infty = \mu_\mathrm{DW} E$, where $\mu_\mathrm{DW} = 1 \cdot 10^{-5} \,\mathrm{m^2 \, s^{-1} V^{-1}}$ is the low-field mobility of the wall and is in good agreement with \eqref{eq:Steady}. This value is computed for the Gilbert damping constant $\alpha = 0.05$, used in \cite{Matsubara15}, and the magnetoelectric coupling constant $\gamma = 1.8 \cdot 10^{-10} \,\mathrm{meV\,m\,V^{-1}}$, estimated from the ferroelectric polarization of a spiral domain in TbMnO$_3$ measured at $10 \, \mathrm{K}$ \cite{Kimura03}.

To analyze how hedgehog defects give rise to intrinsic DW pinning, we simulate a $35 \times 23 \times 46$ system with a single defect line intersecting the meron strings, Fig.~\ref{fig:TopControl}(c,$\,$f). By performing several simulations with different applied electric fields $E$, we observe DW motion only for $E \gtrsim E_\mathrm{pin} \approx 10^7 \,\mathrm{Vm^{-1}}$. The concentration of defects is $\rho_\mathrm{def} = 1/L_z$, as there is one defect line per $L_z$ lattice constants. Therefore, $\rho_\mathrm{def}$ can be controlled by changing the size of the system in the $z$-direction, $L_z$. Simulations with different $L_z$ confirm that the coercive field $E_\mathrm{pin}$ is proportional to the concentration of defects.

\end{document}